\documentclass[a4,12pt]{article}

\usepackage{times}
\usepackage{bm}
\usepackage{graphicx}
\usepackage{color}
\usepackage{verbatim}
\usepackage[colorlinks,citecolor=blue,urlcolor=blue,linkcolor=blue]{hyperref}

\usepackage{lipsum,changepage}

\usepackage{calc}
\newlength{\textwnew}
\usepackage[width=1.085\textwidth,font=small]{caption}

\usepackage{algorithm}
\usepackage[noend]{algpseudocode}
\usepackage{graphicx}
\usepackage[comma,authoryear]{natbib}

\usepackage[below]{placeins}
\usepackage{amsmath,amssymb,booktabs,bm}
\usepackage{amsthm}
\usepackage{rotating}

\newtheorem{theorem}{Theorem}
\newtheorem{definition}{Definition}
\newtheorem{proposition}{Proposition}
\newtheorem{corollary}{Corollary}

\def\bmx{{\boldsymbol x}}
\def\bmX{{\boldsymbol X}}
\def\bmy{{\boldsymbol y}}
\def\bmY{{\boldsymbol Y}}
\def\bmu{{\boldsymbol u}}
\def\bmU{{\boldsymbol U}}
\def\bmv{{\boldsymbol v}}
\def\bmz{{\boldsymbol z}}
\def\bmZ{{\boldsymbol Z}}
\def\bma{{\boldsymbol a}}
\def\bmb{{\boldsymbol b}}
\def\bmr{{\boldsymbol r}}
\def\bmc{{\boldsymbol c}}
\def\bmA{{\boldsymbol A}}
\def\bmm{{\boldsymbol\mu}}
\def\bmS{{\boldsymbol\Sigma}}

\newcommand\argmax{\operatornamewithlimits{argmax}}
\newcommand\conv{\operatornamewithlimits{conv}}
\newcommand\supp{\operatornamewithlimits{supp}}
\newcommand\ave{\operatornamewithlimits{ave}}

\begin{document}

\raggedbottom

\title{Nonparametric imputation by data depth}
\author{Pavlo Mozharovskyi\footnote{
    The major part of this project has been conducted during the postdoc of Pavlo Mozharovskyi at Agrocampus Ouest (Rennes) granted by \textit{Centre Henri Lebesgue} due to program PIA-ANR-11-LABX-0020-01.}\hspace{.2cm}\\
    CREST, Ensai, Universit\'{e} Bretagne Loire\\
    and \\
    Julie Josse \\
    \'{E}cole Polytechnique, CMAP\\
    and \\
    Fran\c{c}ois Husson \\
    IRMAR, Applied Mathematics Unit, Agrocampus Ouest, Rennes\\
    }
\date{August 6, 2018}
\maketitle

\abstract{
We present single imputation method for missing values which borrows the idea of data depth---a measure of centrality defined for an arbitrary point of a space with respect to a probability distribution or  data cloud. This consists in iterative maximization of the depth of each observation with missing values, and can be employed with any properly defined statistical depth function. For each single iteration, imputation reverts to optimization of quadratic, linear, or quasiconcave functions that are solved analytically by linear programming or the Nelder-Mead method. As it accounts for the underlying data topology, the procedure is distribution free, allows imputation close to the data geometry, can make prediction in situations where local imputation ($k$-nearest neighbors, random forest) cannot, and has attractive robustness and asymptotic properties under elliptical symmetry. It is shown that a special case---when using the Mahalanobis depth---has direct connection to well-known methods for the multivariate normal model, such as iterated regression and regularized PCA. The methodology is extended to multiple imputation for data stemming from an elliptically symmetric distribution. Simulation and real data studies show good results compared with existing popular alternatives. The method has been implemented as an \texttt{R}-package. Supplementary materials for the article are available online.
}
\indent\\


\sloppy

\section{Introduction}\label{sec:intro}

Missing data is a ubiquitous problem in statistics. Non-responses to surveys, machines that break and stop reporting, and data that have not been recorded, impede analysis and threaten the validity of inference.
A common strategy \citep{LittleR02} for dealing with missing values is single  imputation, replacing missing entries with plausible values to obtain a completed data set, which can then be analyzed. 

There are two main families of parametric imputation methods: ``joint'' and ``conditional'' modeling, see e.g., \cite{JosseR18} for a literature overview.
Joint modeling specifies a joint distribution for the data, the most popular being the normal multivariate  distribution. The  parameters of the distribution, here the mean and the covariance matrix, are then estimated from the incomplete data using an algorithm such as expectation maximization (EM) \citep{DempsterLR77}. The missing entries are then imputed with the conditional mean, i.e., the conditional expectation of the missing values, given observed values and the estimated parameters.
An alternative is to impute  missing values using a principal component analysis (PCA) model which assumes data are generated as a low rank structure corrupted by Gaussian noise. This method  is closely connected to the literature on matrix completion \cite{JosseH12,HastieMLZ15}, and has shown good imputation capacity due to the plausibility of the low rank assumption \citep{Udell17}. 
The conditional modeling approach \citep{VanBuuren12} consists in specifying one  model for each variable to be imputed, and considers the others variables as explanatory. This procedure is iterated until predictions stabilize. 
Nonparametric imputation methods have also been developed such as imputation by $k$-nearest neighbors ($k$NN) \citep[see][and references therein]{TroyanskayaCSBHTBA01}
or random forest \citep{StekhovenB12}.

\begin{sloppypar}
Most imputation methods are defined under the missing (completely) at random (M(C)AR) assumption, which means that the probability of having missing values does not depend on missing data (nor on observed data).
Gaussian and PCA imputations are sensitive to outliers and deviations from distributional assumptions, whereas nonparametric methods such as  $k$NN  and random forest cannot extrapolate. 
\end{sloppypar}

Here we propose a family of nonparametric imputation methods based on the notion of a statistical depth function \citep{Tukey74}. Data depth is a data-driven multivariate measure of centrality 
that describes data with respect to location, scale, and shape based on a multivariate ordering. It has been applied in multivariate data analysis \citep{LiuPS99}, 
classification \citep{Jornsten04,LangeMM14}, multivariate risk measurement \citep{CascosM07}, and robust linear programming \citep{BazovkinM15}, but has never been applied  in the context of missing data.
Depth based imputation provides excellent predictive properties and has the advantages of both global and local imputation methods.
It imputes close to the data geometry, while still accounting for global features.
In addition, it allows robust imputation in both outliers and heavy-tailed distributions.

\begin{figure}[!t]
\begin{center}
\includegraphics[width=2.825in,trim=10mm 10mm 10mm 17mm,clip]{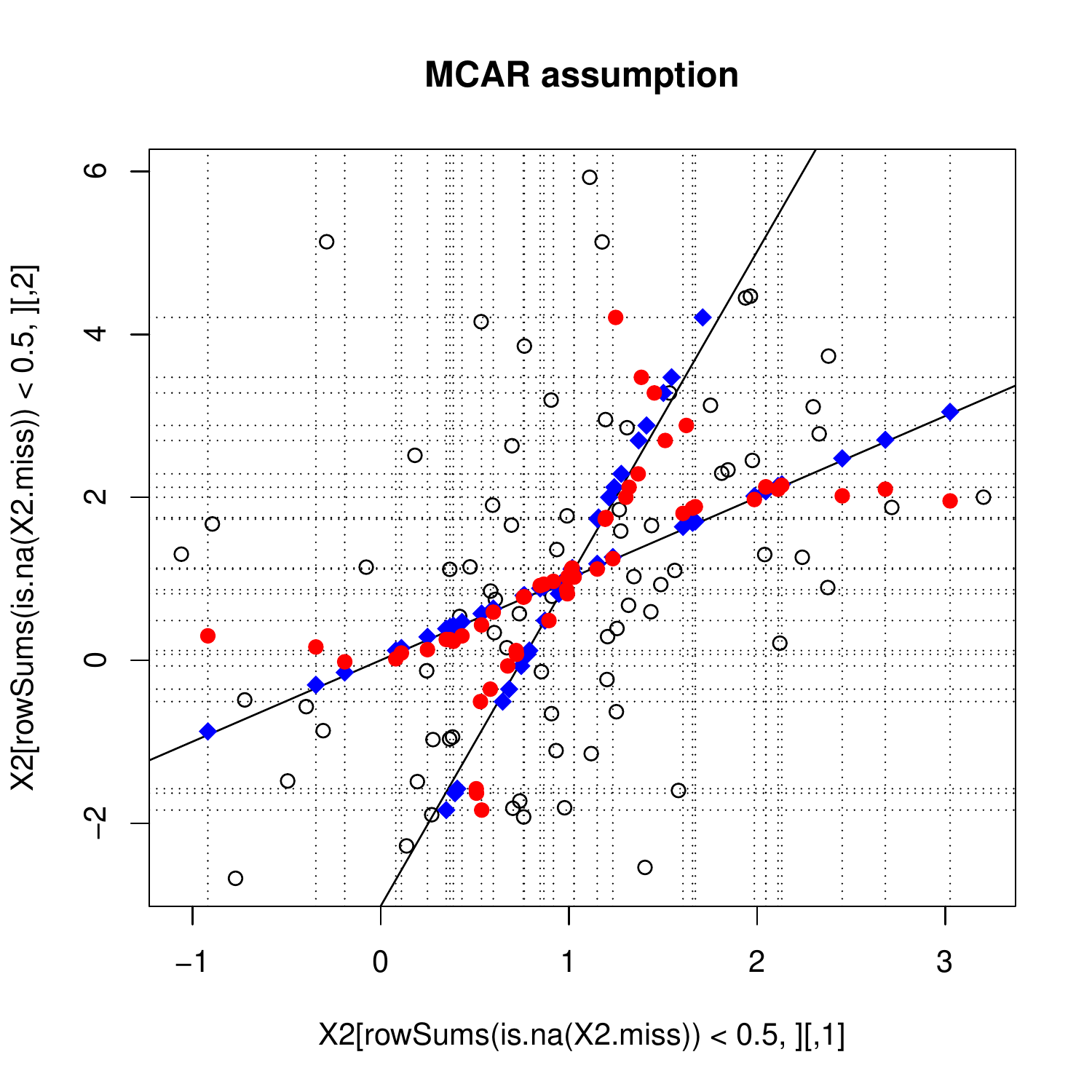}\,\quad\,\includegraphics[width=2.825in,trim=10mm 10mm 10mm 17mm,clip]{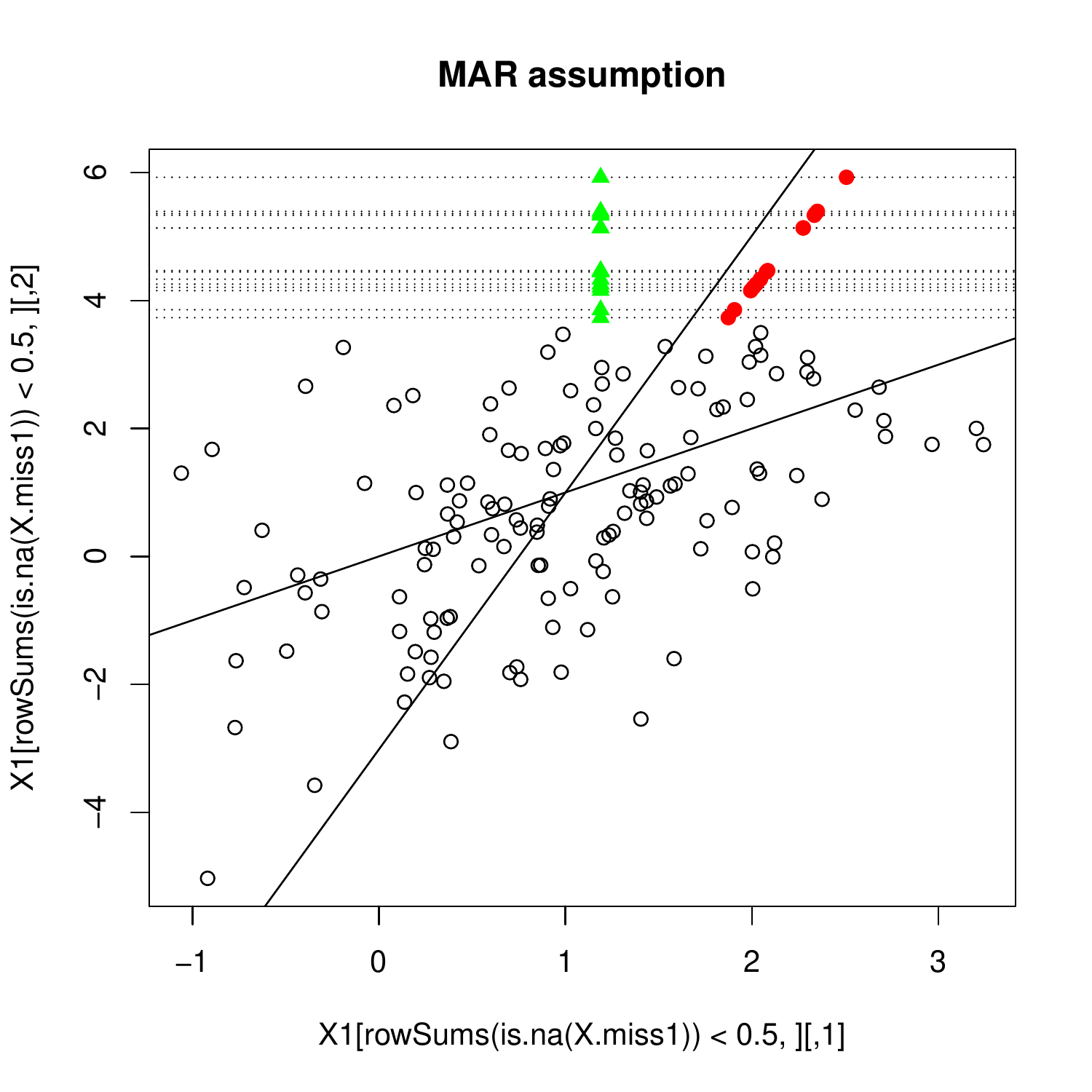}
\end{center}
\caption{Bivariate normal distribution with 30\% MCAR (left) and with MAR in the second coordinate for values $>3.5$ (right); imputation using maximum zonoid depth (filled circles), conditional mean imputation using EM estimates (rhombi), and random forest imputation (triangles). \label{fig:intro12}}
\end{figure}

\begin{figure}[!t]
\begin{center}
\includegraphics[width=2.825in,trim=10mm 10mm 10mm 17mm,clip]{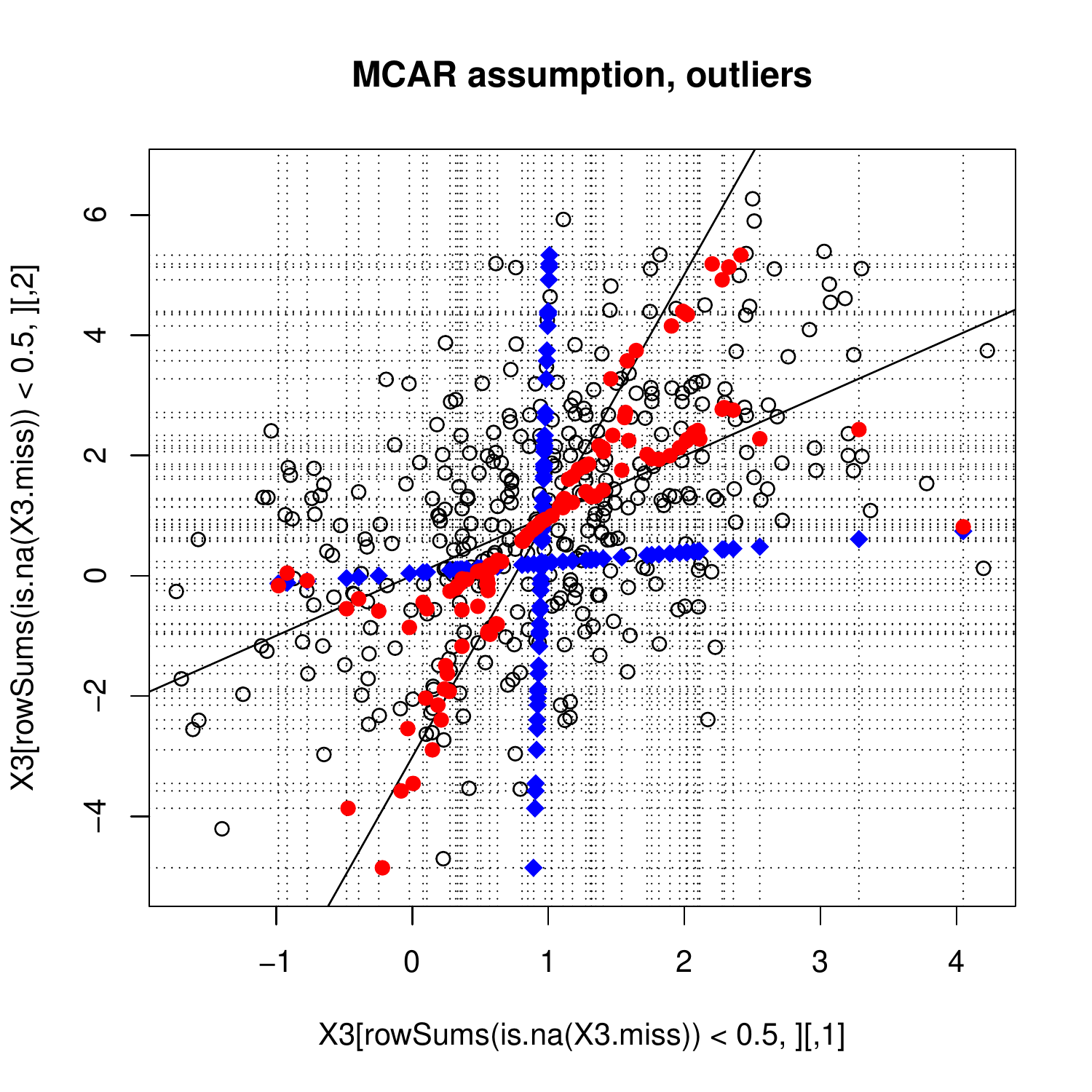}\,\quad\,\includegraphics[width=2.825in,trim=10mm 10mm 10mm 17mm,clip]{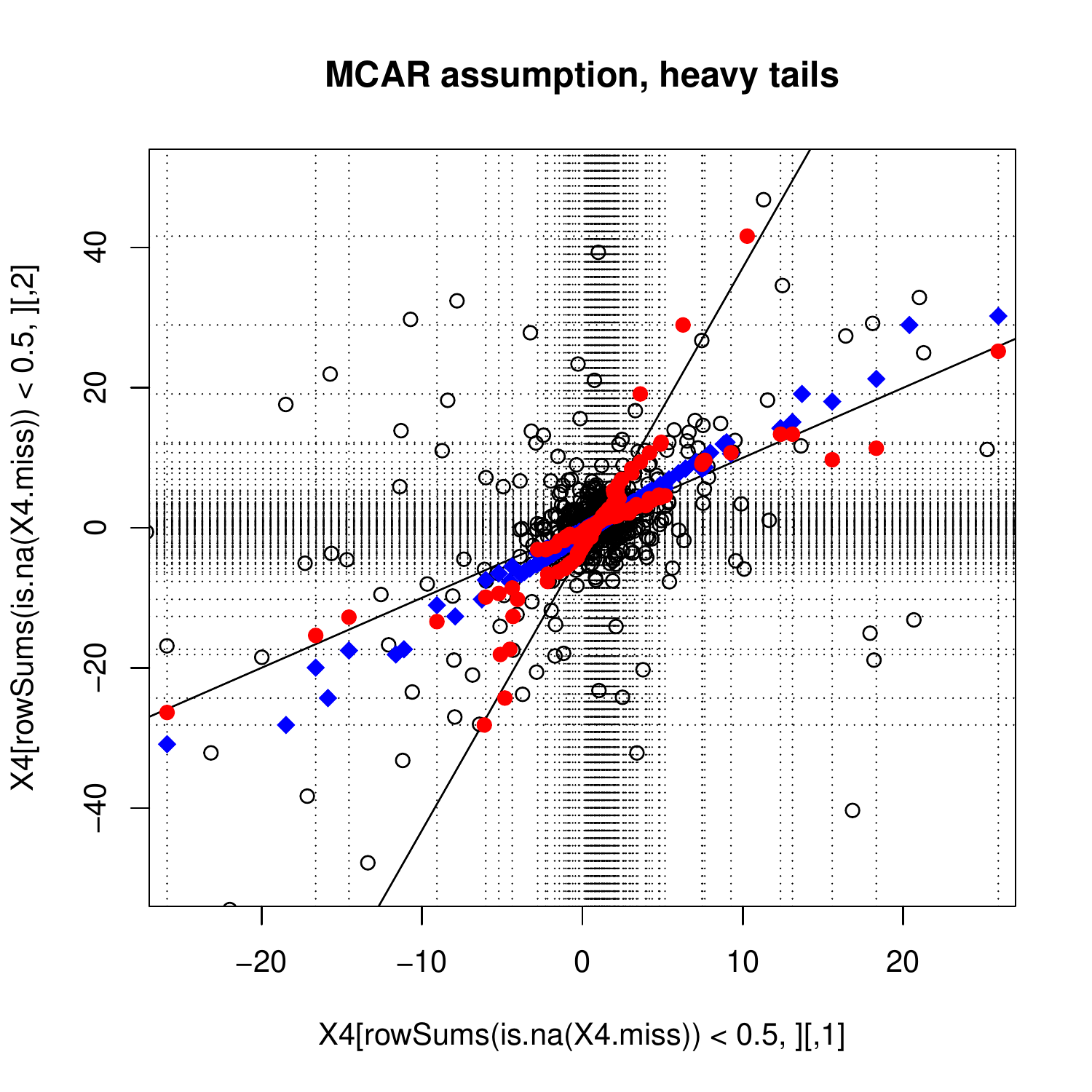}
\end{center}
\caption{Left: Mixture of normal (425 points, 15\% MCAR) and Cauchy (75 points) samples. Right: 1000 Cauchy distributed points with 15\% MCAR. Imputation with Tukey depth (filled circles) and conditional mean imputation using EM estimates (rhombi). \label{fig:intro34}}
\end{figure}

Figures~\ref{fig:intro12} and~\ref{fig:intro34} motivate our proposed depth-based imputation by contrasting it to classical methods.
First, 150 points are drawn from a bivariate normal distribution with mean $\bmm_1=(1,1)^\top$ and covariance $\bmS_1 = \bigl((1, 1)^\top, (1, 4)^\top\bigr)$ and 30\% of the entries are removed completely at random in both variables; points with one missing entry are indicated by dotted lines while solid lines provide (oracle) imputation using distribution parameters. 
The imputation assuming a joint Gaussian distribution using EM estimates is shown by rhombi (Figure~\ref{fig:intro12}, left).
Zonoid depth-based imputation, represented by filled circles, shows that the sample is not necessarily normal, and that this uncertainty increases as we move to the fringes of the data cloud, where imputed points deviate from the conditional mean towards the unconditional one.
Second, the missing values are generated as follows: the first coordinate is removed when the second coordinate $>3.5$ (Figure~\ref{fig:intro12}, right).
Here, the depth-based imputation allows extrapolation when predicting missing values, while the random forest imputation (triangles) gives, as expected, rather poor results.

In Figure~\ref{fig:intro34} (left), we draw 500 points, 425 from the same normal distribution as above, with 15\% of MCAR values and 75 outliers from the Cauchy distribution with the same center and shape matrix and without missing values. In Figure~\ref{fig:intro34} (right), we depict 1000 points drawn from Cauchy distribution  with  15\% MCAR.
As expected, imputation with conditional mean based on EM estimates (rhombi) is rather random. Depth-based imputation with Tukey depth (filled circles) has robust imputed values that are close to the (distribution's) regression lines reflecting data geometry.

The paper is organized as follows. 
Section~\ref{sec:imputation} describes the algorithm for imputing by data depth and derives its theoretical properties under ellipticity.  Section~\ref{sec:whichdepth} describes the special case of imputation with Mahalanobis depth, emphasizing its relationship to existing imputation methods by regression and PCA, and imputation with zonoid and Tukey depths. For each of them, we suggest an efficient optimization strategy. Next, to go beyond ellipticity, we propose imputation with local depth \citep{PaindaveineVB13} appropriate to data with non-convex support. 
Section~\ref{sec:exp} provides a comparative simulation and real data study. Section~\ref{sec:multiple} extends the proposed approach to multiple imputation in order to perform statistical inference with missing values. Section~\ref{sec:conclusions} concludes the article, gathering together some useful remarks. Proofs are available in the supplementary materials.

\section{Imputation by depth maximization}\label{sec:imputation}

\subsection{Imputation by iterative regression} \label{sec:iterativereg}

Let $X$ be a random vector in $\mathbb{R}^d$ and denote $\bmX=(\bmx_1,\ldots,\bmx_n)^\top$ a sample.
For a point $\bmx_i\in\bmX$, we denote $miss(i)$ and $obs(i)$ the sets of its coordinates containing missing and observed values, $|miss(i)|$ and $|obs(i)|$ their corresponding cardinalities. 


Let the rows $\bmx_i$ be i.i.d. draws from $\mathcal{N}({\boldsymbol{\mu}_\bmX}, \boldsymbol{\Sigma}_\bmX)$. One of the simplest conditional methods for imputing missing values consists in the following iterative regression imputation: 
(1) initialize missing values arbitrary, using unconditional mean imputation;
(2) impute missing values in one variable by the values predicted by the regression model of this variable with the remaining variables taken as explanatory ones, 
(3) iterate through variables containing missing values until convergence.
Here, at each step, each point $\bmx_i$ with missing values 
at a coordinate $j$
is imputed with the univariate conditional mean $\mathbb{E}[X|X_{\{1,...,d\}\setminus\{j\}}=\bmx_{i,\{1,...,d\}\setminus\{j\}},\boldsymbol{\mu}_X=\boldsymbol{\mu}_\bmX,\boldsymbol{\Sigma}_X=\boldsymbol{\Sigma}_\bmX]$ with the moment estimates $\bmm_\bmX=\frac{1}{n}\sum_{i=1}^n\bmx_i$ and ${\boldsymbol\Sigma}_\bmX=\frac{1}{n-1}\sum_{i=1}^n(\bmx_i-\bmm_\bmX)(\bmx_i-\bmm_\bmX)^\top$.
After convergence, each point $\bmx_i$ with missing values in $miss(i)$ is imputed with the multivariate conditional mean
\begin{align}\label{equ:condmean}
&\mathbb{E}[X|X_{obs(i)}=\bmx_{i,obs(i)},\boldsymbol{\mu}_X=\boldsymbol{\mu}_\bmX,\boldsymbol{\Sigma}_X=\boldsymbol{\Sigma}_\bmX]\\
\notag =\,&{\boldsymbol\mu}_{\bmX\,miss(i)} + {\boldsymbol\Sigma}_{\bmX\,miss(i),obs(i)}{\boldsymbol\Sigma}^{-1}_{\bmX\,obs(i),obs(i)}\bigl(\bmx_{i,obs(i)} - {\boldsymbol\mu}_{\bmX\,obs(i)}\bigr).
\end{align}
The last expression is the closed-form solution to 
\[
\min_{\bmz_{miss(i)}\in\mathbb{R}^{|miss(i)|}\,,\,\bmz_{obs(i)}=\bmx_{obs(i)}} d_{M}(\bmz,\boldsymbol{\mu}_\bmX|\boldsymbol{\Sigma}_\bmX)
\]
with $d_{M}^2(\bmz,\boldsymbol{\mu}_\bmX|\boldsymbol{\Sigma}_\bmX)=(\bmz-\boldsymbol{\mu}_\bmX)^\top\boldsymbol{\Sigma}_\bmX^{-1}(\bmz-\boldsymbol{\mu}_\bmX)$ being the squared Mahalanobis distance from $\bmz$ to $\boldsymbol{\mu}_\bmX$. Minimizing the Mahalanobis distance can be seen as maximizing a centrality measure---the Mahalanobis depth: 
\[
\max_{\bmz_{miss(i)}\in\mathbb{R}^{|miss(i)|}\,,\,\bmz_{obs(i)}=\bmx_{obs(i)}}D_n^{M}(\bmz|\bmX)\,
\]
where the Manahalobis  depth of $\bmx\in\mathbb{R}^d$ w.r.t. $X$ is defined as follows.
\begin{definition}\emph{\citep{Mahalanobis36}}
$D^{M}(\bmx|X) = \bigl(1 + (\bmx-\bmm_X)^\top{\boldsymbol\Sigma}_X^{-1}(\bmx-\bmm_X)\bigr)^{-1}$,
where $\bmm_X$ and ${\boldsymbol\Sigma}_X$ are the location and shape parameters of $X$.
\end{definition}
In its empirical version $D^{M}_n(\cdot|\bmX)$, the parameters are replaced by their estimates.


The Manahalobis depth is the simplest instance of a statistical depth function. 
We now generalize the iterative imputation algorithm to other depths.

\subsection{Imputation by depth maximization}

\subsubsection{Definition of data depth}


\begin{definition}\emph{\citep{ZuoS00a}}\label{def:depth}
	A bounded non-negative mapping $D(\cdot|X)$ from $\mathbb{R}^d$ to $\mathbb{R}$ is called a \emph{statistical depth function} if it is
\emph{(P1) affine invariant}, i.e., $D(\bmx|X) = D(\bmA\bmx + \bmb|\bmA X + \bmb)$ for any invertible $d\times d$ matrix $\bmA$ and any $\bmb\in\mathbb{R}^d$;
\emph{(P2) maximal at the symmetry center}, i.e., $D(\bmc|X) = \sup_{\bmx\in\mathbb{R}^d}D(\bmx|X)$ for any $X$ halfspace symmetric around $\bmc$ (A random vector $X$ having distribution $P_X$ is said to be halfspace symmetric around (a center) $\bmc\in\mathbb{R}^d$ if $P_X(H)\ge\frac{1}{2}$ for every halfspace $H$ containing $\bmc$.);
\emph{(P3) monotone w.r.t. the deepest point}, i.e., for any $X$ having $\bmc$ as a deepest point, $D(\bmx|X)\le D\bigl(\alpha\bmc + (1 - \alpha\bigr)\bmx)|X)$ for any $\bmx\in\mathbb{R}^d$ and $\alpha\in[0,1]$;
\emph{(P4) vanishing at infinity}, i.e., $\lim_{\|\bmx\|\rightarrow 0}D(\bmx|X) = 0$.
\end{definition}

Additionally, we require (P5) \emph{quasiconcavity} of $D(\cdot|X)$, upper-level sets (or depth-trimmed regions) $D_\alpha(X)=\{\bmx\in\mathbb{R}^d\,:\,D(\bmx|X)\ge\alpha\}$ to be convex, a useful property for optimization. We denote $D_n(\cdot|\bmX)$ the corresponding empirical depth.
See also \cite{ZuoS00b} for a reference on depth contours.

\subsubsection{Imputation by depth maximization} \label{sec:imputedep}

We suggest a unified framework to impute missing values by depth maximization, which extends  iterative regression imputation. More precisely, consider the following iterative scheme:
(1) initialize missing values arbitrarily using unconditional mean imputation;
(2) impute a point $\bmx$ containing missing coordinates with the point $\bmy$ maximizing data depth conditioned on observed values $\bmx_{obs}$:
\begin{equation}\label{equ:impMax}
      \bmy = \argmax_{\bmz_{miss}\in\mathbb{R}^{|miss|}\,,\,\bmz_{obs}=\bmx_{obs}}D_n(\bmz|\bmX)\,;
\end{equation}
(3) iterate until convergence.


The solution of (\ref{equ:impMax}) can be non-unique (see Figure~1 in the supplementary materials for an illustration) and the depth value may become zero immediately beyond the convex hull of the support of the distribution. 
To avoid these problems, we suggest
\emph{imputation by depth} (ID) of an $\bmx$ which has missing values with $\bmy=ID\bigl(\bmx,D_n(\cdot|\bmX)\bigr)$:
\begin{eqnarray}\label{equ:impAve}
	ID\bigl(\bmx,D_n(\cdot|\bmX)\bigr) &=& \ave\bigl(\operatornamewithlimits{arg\,min}_{\bmu\in\mathbb{R}^d\,,\,\bmu_{obs}=\bmx_{obs}}\{\|\bmu - \bmv\|\,|\,\bmv\in D_{n,\alpha^*}(\bmX)\}\bigr),\, \\
	\text{with}\quad\alpha^* &=& \inf_{\alpha\in(0;1)}\bigl\{\alpha\,|\,D_{n,\alpha}(\bmX)\cap\{\bmz\,|\,\bmz\in\mathbb{R}^d\,,\,\bmz_{obs}=\bmx_{obs}\}=\varnothing\bigr\},\,\nonumber
\end{eqnarray}
where $\text{ave}$ is the averaging operator.
The imputation by iterative maximization of depth is summarized in Algorithm~\ref{alg:single}.
The complexity of Algorithm~\ref{alg:single} is $O\bigl(N_\epsilon n_{miss} \Omega(D)\bigr)$. It depends on the data geometry and on the missing values (through the number of  outer-loop iterations $N_\epsilon$ necessary to achieve $\epsilon$-convergence), the number of points containing missing values $n_{miss}$, and the depth-specific complexities for solving \eqref{equ:impAve} $\Omega(D)$ are detailed in subsections of Section~\ref{sec:whichdepth}.



\noindent\hspace*{0\parindent}%
\begin{minipage}{0.875\textwidth}
\begin{algorithm}[H]
\caption{Single imputation}\label{alg:single}
\begin{algorithmic}[1]
\Function{impute.depth.single}{$\boldsymbol{X}$}
\State $\boldsymbol{Y}\gets\boldsymbol{X}$
\State $\boldsymbol{\mu}\gets\hat{\boldsymbol{\mu}}^{(obs)}(\boldsymbol{X})$ \Comment{Calculate mean, ignoring missing values}
\For{$i = 1:n$}
    \If{$miss(i)\,\ne\,\varnothing$}
        \State $\boldsymbol{y}_{i,miss(i)}\gets\boldsymbol{\mu}_{miss(i)}$ \Comment{Impute with unconditional mean}
    \EndIf
\EndFor
\State $I\gets0$
\Repeat \Comment{Iterate until convergence or maximal iteration}
    \State $I\gets I + 1$
    \State $\boldsymbol{Z}\gets\boldsymbol{Y}$
    \For{$i = 1:n$}
        \If{$miss(i)\,\ne\,\varnothing$}
            \State $\boldsymbol{y}_i\gets ID\bigl(\bmy_i,D_n(\cdot|\bmZ)\bigr)$  \Comment{Impute with maximum depth}
        \EndIf
    \EndFor
\Until{$\max_{i\in\{1,...,n\},j\in\{1,...,d\}}|\boldsymbol{y}_{i,j}-\boldsymbol{z}_{i,j}|<\epsilon\,\,\text{\bf or}\,\, I\,=\,I_{max}$}
\State\Return $\boldsymbol{Y}$
\EndFunction
\end{algorithmic}
\end{algorithm}
\end{minipage}

\indent\\

\subsubsection{Theoretical properties for elliptical distributions}

An elliptical distribution is defined as follows (see \cite{FangKN90}, and \cite{LiuS93} in the data depth context).
\begin{definition}\label{def:elliptical}
A random vector $X$ in $\mathbb{R}^d$ is elliptical if and only if there exists a vector $\bmm_X\in\mathbb{R}^d$ and $d\times d$ symmetric and positive semi-definite invertible matrix $\bmS_X=\boldsymbol{\Lambda}\boldsymbol{\Lambda}^\top$ such that for a random vector $U$ uniformly distributed on the unit sphere $\mathcal{S}^{d-1}$ and a non-negative random variable $R$, it holds that $X\overset{D}{=}\bmm_X + R\boldsymbol{\Lambda}U$. We then say that $X\sim\mathcal{E}_d(\bmm_X,\bmS_X,F_R)$, where $F_R$ is the cumulative distribution function of the generating variate $R$.
\end{definition}


Theorem~\ref{thm:onerow}
shows that for an elliptical distribution, imputation of one point with a quasiconcave uniformly consistent depth converges to the center of the conditional distribution when conditioning on the observed values.
Theorem~\ref{thm:onerow} is illustrated in Figure~2 in the supplementary materials.


\begin{theorem}[One row consistency]\label{thm:onerow}
Let $\bmX=(\bmx_1,\ldots,\bmx_n)^\top$ be a data set in $\mathbb{R}^d$ drawn i.i.d. from $X\sim\mathcal{E}_d({\boldsymbol\mu_X},{\boldsymbol\Sigma}_X,F_R)$ with $d\ge 2$, $F_R$ absolutely continuous with strictly decreasing density, and let $\bmx=(\bmx_{obs},\bmx_{miss})\in\mathbb{R}^d$ with $|obs(\bmx)|\ge1$. Further, let $D(\cdot|X)$ satisfy (P1)--(P5) and $D_{n,\alpha}(\bmX)\xrightarrow[n\rightarrow\infty]{a.s.}D_{\alpha}(X)$.
Then for $\bmy=ID\bigl(\bmx,D_n(\cdot|\bmX)\bigr)$,
	\begin{equation*}
		\bigl|\bmy_{miss} - {\boldsymbol\mu}_{X\,miss} - {\boldsymbol\Sigma}_{X\,miss,obs}{\boldsymbol\Sigma}^{-1}_{X\,obs,obs}(\bmx_{obs} - {\boldsymbol\mu}_{X\,obs})\bigr|\xrightarrow[n\rightarrow\infty]{a.s.}0\,.
	\end{equation*}
\end{theorem}


Theorem~\ref{thm:onecolumn} states that if missing values constitute a portion of the sample but are in a single variable, the imputed values converge to the center of the conditional distribution when conditioning on the observed values.

\begin{theorem}[One column consistency]\label{thm:onecolumn}
Let $\bmX=(\bmx_1,\ldots,\bmx_n)^\top$ be a data set in $\mathbb{R}^d$ drawn i.i.d. from $X\sim\mathcal{E}_d({\boldsymbol\mu_X},{\boldsymbol\Sigma}_X,F_R)$ with $d\ge 2$, $F_R$ absolutely continuous with strictly decreasing density, and let $miss(i)=\{j\}$ with probability $p\in(0,1)$ for a fixed $j\in\{1,\ldots,d\}$. Let $D(\cdot|Z)$ satisfy (P1)--(P5) and $D_{n,\alpha}(\bmZ)\xrightarrow[n\rightarrow\infty]{a.s.}D_{\alpha}(Z)$ for $Z=(1 - p)X + pZ^\prime$ with $Z^\prime={\boldsymbol\mu}_{X\,j} - {\boldsymbol\Sigma}_{X\,j,-j}{\boldsymbol\Sigma}^{-1}_{X\,-j,-j}(X_{-j} - {\boldsymbol\mu}_{X\,-j})$.
Further, let $\bmY$ exist such that $\bmy_i=ID\bigl(\bmx_i,D_n(\cdot|\bmY)\bigr)$ if $miss(i)=\{j\}$ and $\bmy_i=\bmx_i$ otherwise. Then, for all $i$ with $miss(i)=\{j\}$ and denoting ${-j}$ for $\{1,...,d\}\setminus\{j\}$,
	\begin{equation*}
		\bigl|\bmy_{i,j} - {\boldsymbol\mu}_{X\,j} - {\boldsymbol\Sigma}_{X\,j,-j}{\boldsymbol\Sigma}^{-1}_{X\,-j,-j}(\bmx_{i,-j} - {\boldsymbol\mu}_{X\,-j})\bigr|\xrightarrow[n\rightarrow\infty]{a.s.}0\,.
	\end{equation*}
\end{theorem}

\section{Which depth to use? } \label{sec:whichdepth}

The generality of the proposed methodology lies in the possibility of using any notion of depth which defines imputation properties.
We focus here on imputation with Manahalobis, zonoid, and Tukey depths. These are of particular interest because they are quasiconcave and require two, one, and zero first moments of the underlying probability measure, respectively.
\begin{corollary}\label{cor:threedepths}
	Theorems~\ref{thm:onerow} and~\ref{thm:onecolumn} hold for the Tukey depth, for the zonoid depth if $\mathbb{E}[\|X\|]<\infty$, and for the Mahalanobis depth if $\mathbb{E}[\|X\|^2]<\infty$.
\end{corollary}

In addition, the function
$f(\bmz_{miss}) = D_n(\bmz|\bmX)$ subject to $\bmz_{obs}=\bmx_{obs}$ in equation \eqref{equ:impMax}, iteratively optimized in Algorithm~\ref{alg:single}, is quadratic for the Mahalanobis depth, continuous inside $\text{conv}(\bmX)$ (the smallest convex set containing $\bmX$) for the zonoid depth, and stepwise discrete for the Tukey depth, which in all cases leads to efficient implementations. 
For a trivariate Gaussian sample, $f(\bmz_{miss})$ is depicted in Figure~1 in the supplementary materials.

The use of a non-quasiconcave depth (e.g., simplicial, spatial \citep{Nagy17}, etc.) results in non-convex optimization when maximizing depth, and this non-stability impedes numerical convergence of the algorithm. 

\subsection{Mahalanobis depth}\label{ssec:depthMahalanobis}

Imputation with the Mahalanobis depth is related to existing methods. First, we show the link with the minimization of the covariance determinant.

\begin{proposition}[Covariance determinant is quadratic in a point's missing entries]\label{prop:reddet}
  Let $\bmX(\bmy)=\bigl(\bmx_1,\ldots,(\bmx_{i,1},\ldots,\bmx_{i,|obs(i)|},\bmy^\top)^\top,\ldots,\bmx_n\bigr)^\top$ be a $n\times d$ matrix with ${\boldsymbol\Sigma}_\bmX(\bmy)$ invertible for all $\bmy\in\mathbb{R}^{|miss(i)|}$. Then $|{\boldsymbol\Sigma}_\bmX(\bmy)|$ is quadratic and globally minimized in $\bmy={\boldsymbol\mu}_{\bmX\,miss(i)}(\bmy) + {\boldsymbol\Sigma}_{\bmX\,miss(i),obs(i)}(\bmy){\boldsymbol\Sigma}_{\bmX\,obs(i),obs(i)}^{-1}(\bmy)\bigl((\bmx_{i,1},\ldots,\bmx_{i,|obs(i)|}) - {\boldsymbol\mu}_{\bmX\,obs(i)}\bigr)$.
\end{proposition}

From Proposition~\ref{prop:reddet} it follows  that the minimum of the covariance determinant is unique and the determinant itself decreases at each iteration.
Thus, to impute points with missing coordinates one-by-one and iterate until convergence constitutes the block coordinate descent method, which can be proved to numerically converge due to Proposition~2.7.1 from~\cite{Bertsekas99} (as long as $\boldsymbol{\Sigma}_\bmX$ is invertible).



Further, Theorem \ref{thm:Mahalanobis} states that imputation using the maximum Mahalanobis depth, iterative (multiple-output) regression, and regularized PCA \citep{JosseH12} with $S=d-1$ dimensions, all converge to the same imputed sample.

\begin{theorem}\label{thm:Mahalanobis}
Suppose that  we impute $\bmX=(\bmX_{miss},\bmX_{obs})$ in $\mathbb{R}^d$ with $\bmY$ so that $\bmy_i = \argmax_{\bmz_{obs(i)}=\bmy_{obs(i)}}\,\,\,D^{M}_n(\bmz|\bmY)$ for each $i$ with $|miss(i)|>0$ and $\bmy_i=\bmx_i$ otherwise.
 %
  Then for each such $\bmy_i$,
   it also holds that:
  \begin{itemize}
    \item $\bmx_{i}$ is imputed with the {\bf conditional mean}:
\begin{equation*}
      \bmy_{i,miss(i)} = {\boldsymbol\mu}_{\bmY\,miss(i)} + {\boldsymbol\Sigma}_{\bmY\,miss(i),obs(i)}{\boldsymbol\Sigma}^{-1}_{\bmY\,obs(i),obs(i)}(\bmx_{obs(i)} - {\boldsymbol\mu}_{\bmY\,obs(i)})\,
\end{equation*}
     which is equivalent to {\bf single}- and {\bf multiple}-output {\bf regression},
    \item $\bmY$ is a {\bf stationary point} of $|\bmS_\bmX(\bmX_{miss})|$:
    $\frac{\partial|\bmS_\bmX|}{\partial\bmX_{miss}}(\bmY_{miss}) = \boldsymbol{0},$ and
    \item each missing coordinate $j$ of $\bmx_i$ is imputed with {\bf regularized PCA} as in  Josse \& Husson (2012) with any $0<\sigma^2\le\lambda_d$ and with $\bmX-{\boldsymbol\mu}_\bmX={\boldsymbol U}{\boldsymbol\Lambda}^\frac{1}{2}{\boldsymbol V}^\top$  the singular value decomposition (SVD):
      $\bmy_{i,j} = \sum_{s=1}^d{\boldsymbol U}_{i,s}\sqrt{\frac{\lambda_s-\sigma^2}{\lambda_s}}{\boldsymbol V}_{j,s} + \bmm_{\bmY\,j}\,.$

  \end{itemize}
\end{theorem}

The first point of the theorem sheds light on the connection between imputation by Mahalanobis depth and the iterative regression imputation of Section \ref{sec:iterativereg}. 
When the Mahalanobis depth is used in Algorithm~\ref{alg:single}, each $\bmx_i$ with missingness in $ miss(i)$ is imputed by the multivariate conditional mean as in equation \eqref{equ:condmean}, and thus lies  in the $\bigl(d-|miss(i)|\bigr)$-dimensional multiple-output regression subspace of $\bmX_{\cdot,miss(i)}$ on $\bmX_{\cdot,obs(i)}$. This subspace is obtained as the intersection of the single-output regression hyperplanes $\bmX_{\cdot,j}$ on $\bmX_{\cdot, \{1,...,d\}\setminus\{j\}}$ for all $j\in miss(i)$  corresponding to missing coordinates.
The third point  strengthens the method as imputation with regularized PCA has proved to be highly efficient in practice due to its sticking to  low-rank structure of importance and ignoring noise.

The complexity of imputing a single point with the Mahalanobis depth is $O(nd^2 + d^3)$. Despite its good properties, it is not robust to outliers. However, robust estimates for ${\boldsymbol\mu}_\bmX$ and ${\boldsymbol\Sigma}_\bmX$  can be used, e.g., the minimum covariance determinant ones \citep[MCD, see][]{RousseeuwVD99}.

\subsection{Zonoid depth}

\cite{KoshevoyM97} define a zonoid trimmed region, with $\alpha\in (0,1]$, as
\begin{equation*}
  D^z_{\alpha}(X)=\Bigl\{\int_{{\mathbb R}^d}\bmx g(\bmx)dP_X(\bmx):\,g\,:\,{\mathbb R}^d\mapsto\left[0,\frac{1}{\alpha}\right]\,\mbox{measurable and}\,\int_{{\mathbb R}^d}g(\bmx)dP_X(\bmx)=1\Bigr\}
\end{equation*}
and for $\alpha=0$ as $D^z_0(X)=\mbox{cl}\left(\cup_{\alpha\in(0,1]}D^z_\alpha(X)\right)$,
where $\mbox{cl}$ denotes the closure.
Its empirical version can be defined as
\begin{equation*}
  D^{z}_{n,\alpha}(\bmX)=\Bigl\{\sum_{i=1}^n\lambda_i\bmx_i\,:\,\sum_{i=1}^n\lambda_i=1\,,\,\lambda_i\ge 0\,,\,\alpha\lambda_i\le\frac{1}{n}\,\,\forall\,\,i\in\{1,\ldots,n\}\Bigr\}\,.
\end{equation*}
\begin{definition}\emph{\citep{KoshevoyM97}}
  The zonoid depth of $\bmx$ w.r.t. $X$ is defined as
  \begin{equation*}
    D^z(\bmx|X)=\left\{
    \begin{array}{l l}
        \sup\{\alpha:\,\bmx\in D^z_{\alpha}(X)\} & \quad \text{if $\bmx\in\conv\bigl(\supp(X)\bigr)$},\\
        0                                             & \quad \text{otherwise}.
    \end{array}\right.\
\end{equation*}
\end{definition}
For a comprehensive reference on the zonoid depth, the reader is referred to~\cite{Mosler02}.


Imputation of a point $\bmx_i$ in Algorithm 1 is then performed  by a slight modification of the linear programming for computation of zonoid depth  with variables $\gamma$ and $\boldsymbol{\lambda}=(\lambda_1,...,\lambda_n)^\top$:
\begin{equation}\label{eqn:linprog}
  \min\,\gamma\quad\text{s.t.}\quad
  \bmX_{\cdot,obs(i)}^\top\boldsymbol{\lambda} = \bmx_{i,obs(i)}\,,
  \notag\boldsymbol{\lambda}^\top\boldsymbol{1}_n = 1\,,
  \notag\gamma\boldsymbol{1}_n - \boldsymbol{\lambda} \ge \boldsymbol{0}_n\,,
  \notag\boldsymbol{\lambda} \ge \boldsymbol{0}_n\,.
\end{equation}
Here $\bmX_{\cdot,obs(i)}$ stands for the completed $n\times|obs(i)|$ data matrix containing columns corresponding only to non-missing coordinates of $\bmx_i$, and $\boldsymbol{1}_n$ (respectively $\boldsymbol{0}_n$) is a vector of ones (respectively zeros) of length $n$. In the implementation, we use the simplex method, which is known for being fast despite its exponential complexity. This implies that, for each point $\bmx_i$, imputation is performed by the weighted mean:
\begin{equation*}
  \bmy_{i,miss(i)} = \bmX^\top_{\cdot,miss(i)}\boldsymbol{\lambda}\,,
\end{equation*}
the average of the maximum number of equally weighted points.
Additional insight on the position of imputed points with respect to the sample can be gained by inspecting the optimal weights $\lambda_i$. 
Zonoid imputation is related to local methods such as as $k$NN imputation, as only some of the weights are 
positive.


\subsection{Tukey depth}

\begin{definition}\emph{\citep{Tukey74}}
  The Tukey depth of $\bmx$ w.r.t. $X$ is defined as
    $D^T(\bmx|X) = \inf\{P_X(H)\,:\,H\text{ a closed halfspace},\,\bmx\in H\}$.
\end{definition}

In the empirical version, the probability is substituted by the portion of $\bmX$ giving
    $D_n^T(\bmx|\bmX) = \min_{\bmu\in\mathcal{S}^{d-1}}\frac{1}{n}\bigl|\{i:\bmx_i^\top\bmu \ge \bmx^\top\bmu\,,\,i=1,...,n\}\bigr|$.
For more information on Tukey depth see \cite{DonohoG92}.

With nonparametric imputation by Tukey depth, one can expect that after convergence of Algorithm~\ref{alg:single}, for each point initially containing missing values, it holds that $\bmy_i = \argmax_{\bmz_{obs}=\bmx_{obs}} \min_{\bmu\in \mathcal{S}^{d-1}}\bigl|\bigl\{k\,:\,\bmy_k^\top\bmu\ge\bmz^\top\bmu,\,k\in\{1,...,n\}\bigr\}\bigr|$. Thus, imputation is performed according to the maximin principle based on criteria involving indicator functions, which implies robustness of the solution.
Note that as the Tukey depth is not continuous, the searched-for maximum \eqref{equ:impMax} may be non-unique (see Figure~1 (top right) in the supplementary materials), and we impute with the barycenter of the maximizing arguments \eqref{equ:impAve}.
Due to the combinatorial nature of the Tukey depth, to speed up implementation, we run $2d$  times the Nelder-Mead downhill-simplex algorithm, and take the average over the solutions. 
The imputation is illustrated in Figure~3 in the supplementary materials.



The Tukey depth can be computed exactly \citep{DyckerhoffM16} with complexity $O(n^{d-1}\log n)$, although to avoid computational burden we also implement its approximation with random directions \citep{Dyckerhoff04} having complexity $O(kn)$, with $k$ denoting the number of random directions. All of the experiments are performed with exactly computed Tukey depth, unless stated otherwise.


\subsection{Beyond ellipticity: local depth}\label{ssec:depthLocal}

Imputation with the so-called ``global depth'' from Definition~\ref{def:depth} may be appropriate in applications even if the data moderately deviate from ellipticity (see Section~\ref{ssec:skewednc}). However, it can fail 
when the distribution has non-convex support or  several modes. A solution is to use the local depth in Algorithm~\ref{alg:single}.

\begin{definition}\emph{\citep{PaindaveineVB13}}
	For a depth $D(\cdot|X)$, the $\beta$-local depth is defined as
		$LD^\beta(\cdot,X)\,:\,\mathbb{R}^d\rightarrow\mathbb{R}^+\,:\,\bmx\mapsto LD^\beta(\bmx,X)=D(\bmx|X^{\beta,\bmx})$ with $X^{\beta,\bmx}$  the conditional distribution of $X$ conditioned on $\bigcap_{\alpha\ge 0,\,P_Y(D_\alpha(Y))\ge\beta}D_\alpha(Y)$, where $Y$ has the distribution $P_Y=\frac{1}{2}P_X + \frac{1}{2}P_{2\bmx - X}$.
\end{definition}
The locality level $\beta$ should be chosen in a data-driven way, for instance by cross-validation. An important advantage of this approach is that any depth satisfying Definition~\ref{def:depth} can be plugged in to the local depth.
We suggest using the Nelder-Mead algorithm to enable imputation with maximum local depth regardless of the chosen depth notion.







\subsection{Dealing with outsiders}

A number of depths that exploit the geometry of the data are equal to zero beyond $\text{conv}(\bmX)$, including the zonoid and Tukey depths.
Although (\ref{equ:impAve}) deals with this situation, for a finite sample it means that points with missing values having the maximal value in at least one of the observed coordinates will never move from the initial imputation because they will become vertices of the $\text{conv}(\bmX)$. For the same reason, other points to be imputed and lying exactly on the $\text{conv}(\bmX)$ will not move much during imputation iterations. As such points are not numerous and would need  to move quite substantially to influence  imputation quality, we impute them---during the initial iterations---using the spatial depth function \citep{VardiZ00}, which is everywhere non-negative. This resembles the so-called ``outsider treatment'' introduced by~\cite{LangeMM14}. Another possibility is to extend the depth beyond $\text{conv}(\bmX)$, see e.g., \cite{EinmahlLL15} for the Tukey depth.

\section{Experimental study}\label{sec:exp}


\subsection{Choice of competitors}\label{ssec:expCompetitors}

We assess the prediction abilities of Tukey, zonoid, and Mahalanobis depth imputation, and the robust Mahalanobis depth imputation using MCD mean and covariance estimates, with the robustness parameter chosen in an optimal way due to knowledge of the simulation setting.  We measure their performance against the competitors: conditional mean imputation based on EM estimates of the mean and covariance matrix; regularized PCA imputation with rank 1 and 2; 
  two nonparametric imputation methods: random forest (using the default implementation in the \texttt{R}-package \texttt{missForest}), and $k$NN imputation 
choosing $k$ from $\{1,\ldots,15\}$, minimizing the imputation error over 10 validation sets as in~\citet{StekhovenB12}. Mean and oracle (if possible) imputations are used to benchmark the results.

\subsection{Simulated data}\label{ssec:expSimulation}

\subsubsection{Elliptical setting with Student-$t$ distribution}

We generate $100$ points according to an elliptical distribution (Definition~\ref{def:elliptical}) with $\bmm_2=(1, 1, 1)^\top$ and the shape $\bmS_2=\bigl((1, 1, 1)^\top, (1, 4, 4)^\top, (1, 4, 8)^\top\bigr)$, where $F$ is the univariate Student-$t$ distribution ranging in number of degrees of freedom (d.f.) from the Gaussian to the Cauchy: $t=\infty,10,5,3,2,1$. For each of the $1000$ simulations, we remove $5\%$, $15\%$ and $25\%$ of values completely at random (MCAR), and compute the median and the median absolute deviation from the median (MAD) of the root mean square error (RMSE) of each imputation method.  
 Table~\ref{tab:StudentT100} presents the results for $25\%$ missing values. The conclusions with other percentages (see the supplementary materials) are the same, but as expected, performances decrease with increasing percentage of missing data.
\begin{table}[!ht]
  \begin{center}
  {\tiny
    \begin{tabular}{|c|c|c|c|c|c|c|c|c|c|c|c|}
    \hline
Distr. & $D^{Tuk}$ & $D^{zon}$ & $D^{Mah}$ & $D^{Mah}_{MCD.75}$ & EM & regPCA1 & regPCA2 & $k$NN & RF & mean & oracle \\
    \hline
$t\,\infty$ & 1.675 & {\it 1.609} & 1.613 & 1.991 & {\bf 1.575} & 1.65 & 1.613 & 1.732 & 1.763 & 2.053 & 1.536 \\
 & (0.205)  & ({\it 0.1893}) & (0.1851) & (0.291) & ({\bf 0.1766}) & (0.1846) & (0.1856) & (0.2066) & (0.2101) & (0.2345) & (0.1772) \\ \cline{1-1}
$t\,10$ & 1.871 & 1.81 & {\it 1.801} & 2.214 & {\bf 1.755} & 1.836 & {\it 1.801} & 1.923 & 1.96 & 2.292 & 1.703 \\
 & (0.2445) & (0.2395) & ({\it 0.2439}) & (0.3467) & ({\bf 0.2379}) & (0.2512) & ({\it 0.2433}) & (0.2647) & (0.2759) & (0.2936) & (0.2206) \\ \cline{1-1}
$t\,5$ & 2.143 & 2.089 & {\it 2.079} & 2.462 & {\bf 2.026} & 2.108 & {\it 2.08} & 2.235 & 2.259 & 2.612 & 1.949 \\
 & (0.3313) & (0.3331) & ({\it 0.3306}) & (0.4323) & ({\bf 0.3144}) & (0.3431) & ({\it 0.3307}) & (0.3812) & (0.3656) & (0.3896) & (0.3044) \\ \cline{1-1}
$t\,3$ & 2.636 & 2.603 & 2.62 & 2.946 & {\bf 2.516} & {\it 2.593} & 2.619 & 2.757 & 2.79 & 3.165 & 2.384 \\
 & (0.5775) & (0.5774) & (0.5745) & (0.6575) & ({\bf 0.5537}) & ({\it 0.561}) & (0.5741) & (0.5874) & (0.5856) & (0.6042) & (0.5214) \\ \cline{1-1}
$t\,2$ & {\bf 3.563} & 3.73 & 3.738 & 3.989 & {\it 3.567} & 3.692 & 3.738 & 3.798 & 3.849 & 4.341 & 3.175 \\
 & ({\bf 1.09}) & (1.236) & (1.183) & (1.287) & ({\it 1.146}) & (1.186) & (1.19) & (1.133) & (1.19) & (1.252) & (0.9555) \\ \cline{1-1}
$t\,1$ & {\it 16.58} & 19.48 & 19.64 & {\bf 16.03} & 18.5 & 18.22 & 19.61 & 17.59 & 17.48 & 20.32 & 13.55 \\
 & ({\it 13.71}) & (16.03) & (16.2) & ({\bf 12.4}) & (15.46) & (15.02) & (16.1) & (14.59) & (14.33) & (16.36) & (10.71) \\ \hline
    \end{tabular}
  }
  \caption{Median and MAD of the RMSEs of the imputation for a sample of $100$ points drawn from  elliptically symmetric Student-$t$ distributions with  $\bmm_2$ and $\bmS_2$  with $25\%$ of  MCAR values, over 1000 repetitions. Bold values indicate the best results, italics the second best.}
  \label{tab:StudentT100}
  \end{center}
\end{table}
As expected, the behavior of the different imputation methods changes with the number of d.f., as does the overall leadership trend. For the Cauchy distribution, robust methods perform best:  Mahalanobis depth-based imputation using MCD estimates, followed closely  by the one using Tukey depth. For $2$ d.f., when the first moment exists but not the second, EM- and Tukey-depth-based imputations perform similarly, with a slight advantage to the Tukey depth in terms of  MAD. For larger numbers of d.f., when two first moments exist, EM takes the lead. It is followed by the group of  regularized PCA methods, and Mahalanobis- and zonoid-depth-based imputation. Note that the Mahalanobis depth and regularized PCA with rank two perform similarly (the small difference can be explained by numerical precision considerations), see Theorem~\ref{thm:Mahalanobis}. Both nonparametric methods perform poorly, being ``unaware'' of the ellipticity of the underlying distribution, but give reasonable results for the Cauchy distribution because of insensitivity to correlation. By default, we present the results obtained with spatial depth for the outsiders. For the Tukey depth, implementation is also available using the extension by \cite{EinmahlLL15}.

\subsubsection{Contaminated elliptical setting}\label{ssec:StudentOutl}

We then modify the above setting by adding $15\%$ of outliers (which do not contain missing values) that stem from the Cauchy distribution with the same parameters $\bmm_2$ and $\bmS_2$. 
\begin{table}[!ht]
  \begin{center}
  {\tiny
    \begin{tabular}{|c|c|c|c|c|c|c|c|c|c|c|c|}
    \hline
Distr. & $D^{Tuk}$ & $D^{zon}$ & $D^{Mah}$ & $D^{Mah}_{MCD.75}$ & EM & regPCA1 & regPCA2 & $k$NN & RF & mean & oracle \\
    \hline
$t\,\infty$ & {\bf 1.751} & 1.86 & 1.945 & {\it 1.81} & 1.896 & 1.958 & 1.945 & 1.859 & 1.86 & 2.23 & 1.563 \\
 & ({\bf 0.2317}) & (0.3181) & (0.4299) & ({\it 0.239}) & (0.3987) & (0.4495) & (0.4328) & (0.2602) & (0.2332) & (0.3304) & (0.1849) \\ \cline{1-1}
$t\,10$ & {\bf 1.942} & 2.087 & 2.165 & {\it 2.022} & 2.112 & 2.196 & 2.165 & 2.051 & 2.047 & 2.48 & 1.733 \\
 & ({\bf 0.2976}) & (0.4295) & (0.5473) & ({\it 0.3128}) & (0.5226) & (0.5729) & (0.5479) & (0.3143) & (0.3043) & (0.4163) & (0.2266) \\ \cline{1-1}
$t\,5$ & {\bf 2.178} & 2.333 & 2.421 & {\it 2.231} & 2.376 & 2.398 & 2.421 & 2.315 & 2.325 & 2.766 & 1.939 \\
 & ({\bf 0.3556}) & (0.4924) & (0.6026) & ({\it 0.381}) & (0.5715) & (0.6035) & (0.5985) & (0.3809) & (0.3946) & (0.528) & (0.2979) \\ \cline{1-1}
$t\,3$ & {\bf 2.635} & 2.864 & 2.935 & {\it 2.664} & 2.828 & 2.916 & 2.93 & 2.797 & 2.838 & 3.34 & 2.356 \\
 & ({\bf 0.6029}) & (0.7819) & (0.8393) & ({\it 0.5877}) & (0.7773) & (0.8221) & (0.8384) & (0.6045) & (0.6228) & (0.7721) & (0.4946) \\ \cline{1-1}
$t\,2$ & {\bf 3.763} & 4.082 & 4.136 & {\it 3.783} & 4.036 & 4.09 & 4.14 & 3.955 & 4.026 & 4.623 & 3.323 \\
 & ({\bf 1.17}) & (1.535) & (1.501) & ({\it 1.224}) & (1.518) & (1.585) & (1.503) & (1.265) & (1.354) & (1.561) & (1.04) \\ \cline{1-1}
$t\,1$ & {\it 17.17} & 20.43 & 20.27 & {\bf 16.46} & 19.01 & 19.81 & 20.53 & 18.96 & 19.04 & 21.04 & 14.44 \\
 & ({\it 13.27}) & (15.99) & (15.91) & ({\bf 12.94}) & (15.21) & (16.15) & (16.28) & (14.73) & (14.62) & (15.56) & (11.33) \\ \hline
    \end{tabular}
  }
  \caption{Median and MAD of the RMSEs of the imputation for $100$ points drawn from  elliptically symmetric Student-$t$ distributions, with $\bmm_2$ and $\bmS_2$ contaminated with $15\%$ of outliers, and $25\%$ of MCAR values on non-contaminated data, repeated 1000 times.}
  \label{tab:StudentOutlT100}
  \end{center}
\end{table}
As expected, Table~\ref{tab:StudentOutlT100} shows that the best RMSEs are obtained by the robust imputation methods: Tukey depth and Mahalanobis depth with MCD estimates. Being restricted to a neighborhood, nonparametric methods often impute based on non-outlying points, and thus perform less well as the preceding group. The rest of the included imputation methods cannot deal with the  contaminated data and perform rather poorly.

\subsubsection{The MAR setting}

\begin{figure}[!h]
  \begin{center}
    \includegraphics[keepaspectratio=true,scale=0.3,trim=0mm 15mm 0mm 15mm,clip=true]{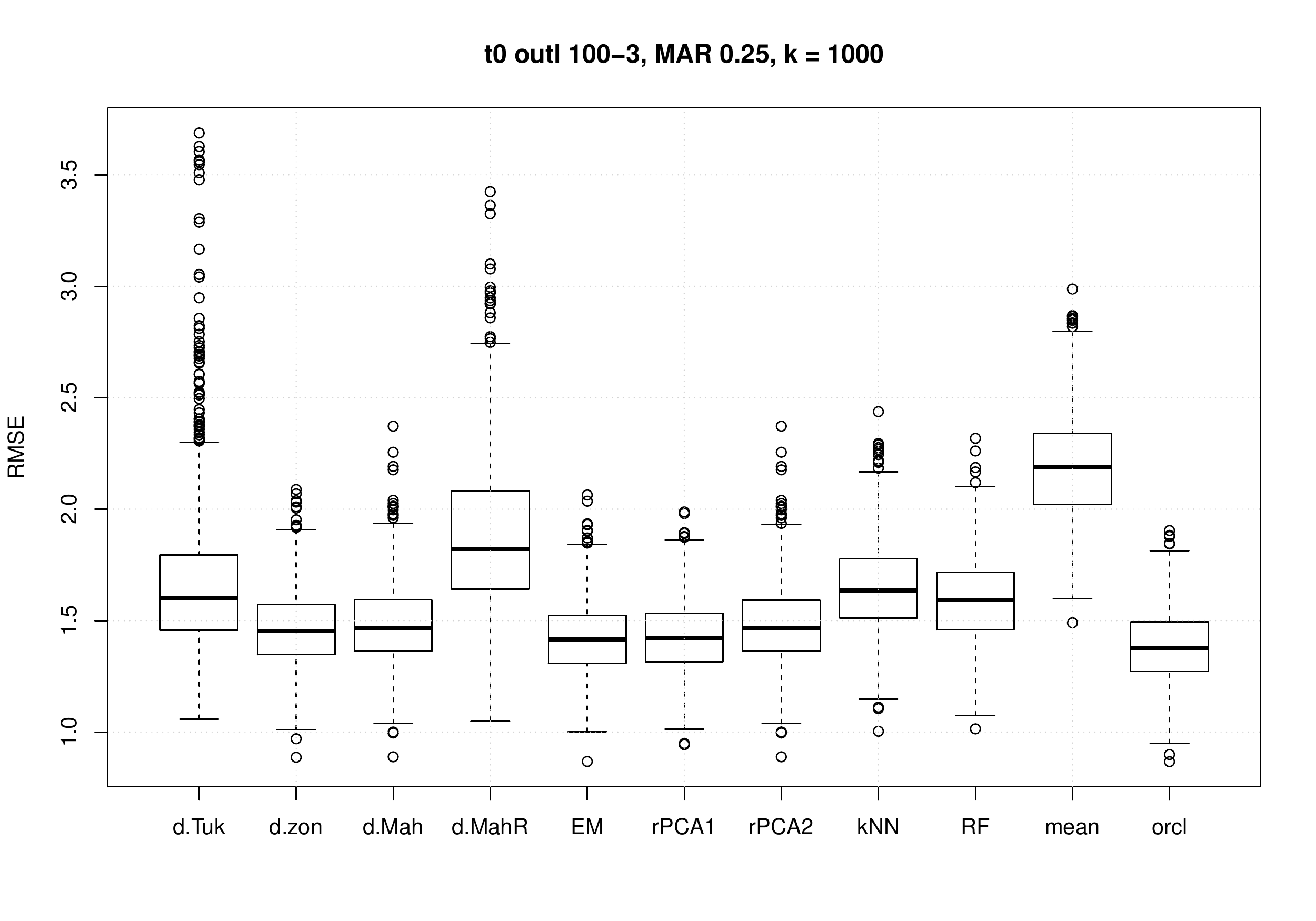}\,\includegraphics[keepaspectratio=true,scale=0.3,trim=10mm 15mm 0mm 15mm,clip=true]{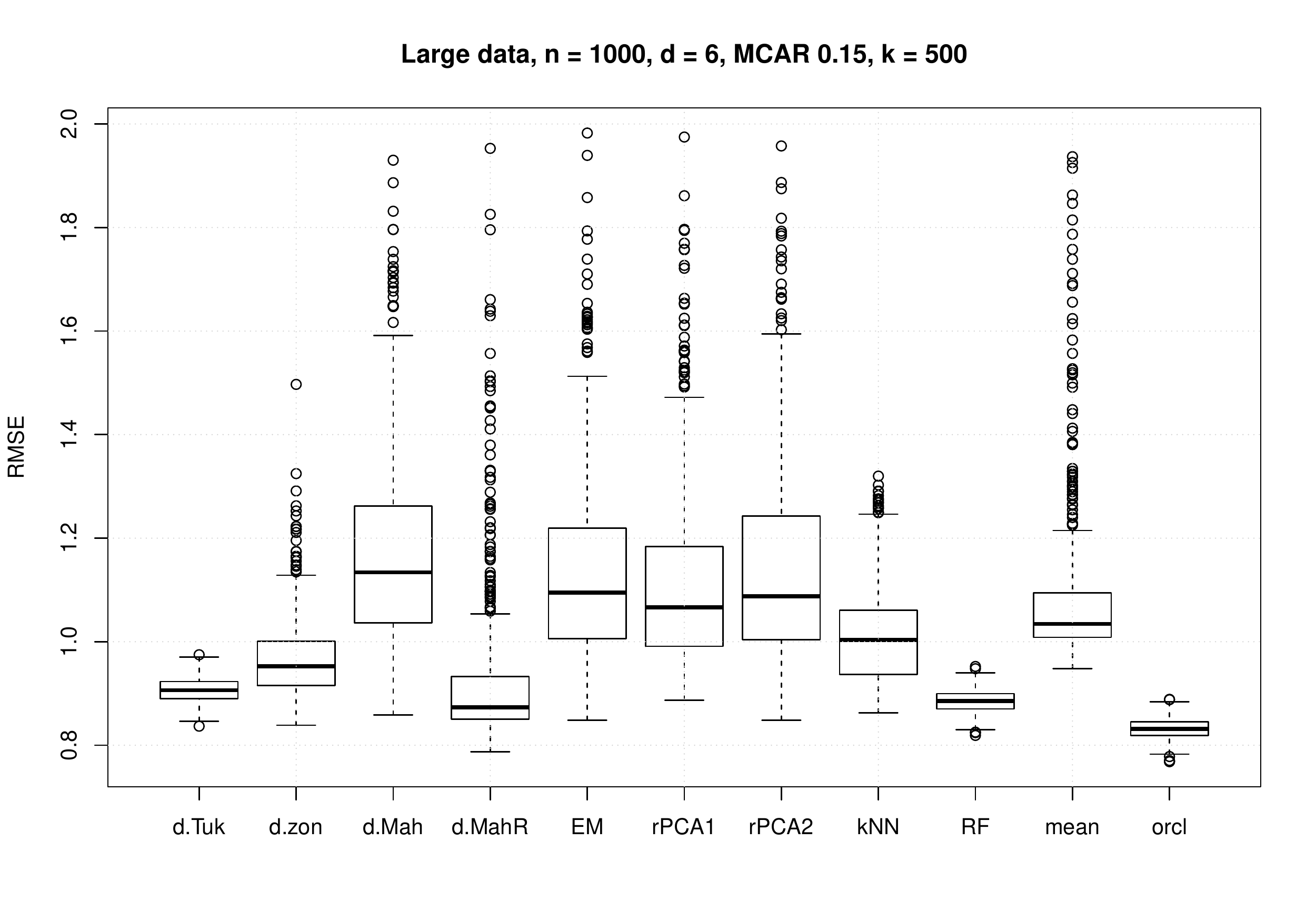}
  \caption{Left: RMSEs for different imputation methods for $100$ points drawn from a correlated $3$-dimensional Gaussian distribution with $\bmm_3$ and $\bmS_3$ with MAR values (see implementation for details), over $1000$ repetitions. Right: $1000$ points drawn from a $6$-dimensional Gaussian distribution with $\bmm_4$ and $\bmS_4$ contaminated with $15\%$ of outliers, and $15\%$ of MCAR values on non-contaminated data, over $500$ repetitions.}
  \label{fig:mar}
  \end{center}
\end{figure}

We next generate highly correlated Gaussian data by setting $\bmm_3=(1, 1, 1)$ and the covariance matrix to $\bmS_3=\bigl((1, 1.75, 2)^\top, (1.75, 4, 4)^\top, (2, 4, 8)^\top\bigr)$. We insert missing values according to the MAR mechanism: the first and third variables are missing depending on the value of the second variable.
Figure~\ref{fig:mar} (left) shows the boxplots of the RMSEs.
As we expected, semiparametric methods (EM, regularized PCA and Mahalanobis depth) perform close to the oracle imputation. The good performance of the rank 1 regularized PCA can be explained by the high correlation between variables. The zonoid depth imputes well despite having no parametric knowledge.
Nonparametric methods are unable to capture the correlation, while robust methods ``throw away'' points possibly containing valuable information.

\subsubsection{The low-rank model}

We consider as a stress-test  an extremely contaminated low-rank model by adding to a 2-dimensional low-rank structure a Cauchy-distributed noise.
Generally, while capturing any structure is rather meaningless in this setting (confirmed by the high MADs in  Table~\ref{tab:lowrank50}), the performance of the methods is ``proportional to the way they ignore'' dependency information. For this reason, mean imputation as well as nonparametric methods perform best. The Tukey and zonoid depths perform second best by accounting only for fundamental features of the data. This can be also said about the regularized PCA when keeping the first principal component only. The remaining methods try to reflect the data structure, but are distracted either by the low rank or  the heavy-tailed noise.


\begin{table}[!h]
  \begin{center}
  {\scriptsize
\begin{tabular}{|l||c|c|c|c|c|c|c|c|c|c|c|}
    \hline
 & $D^{Tuk}$ & $D^{zon}$ & $D^{Mah}$ & $D^{Mah}_{MCD.75}$ & EM & regPCA1 & regPCA2 & $k$NN & RF & mean \\ \hline
Median RMSE &  0.4511 & 0.4536 & 0.4795 & 0.5621 & 0.4709 & 0.4533 & 0.4664 & {\bf 0.4409} & 0.4444 & {\it 0.4430} \\
Mad of RMSE &  0.3313 & 0.3411 & 0.3628 & 0.4355 & 0.3595 & 0.3461 & 0.3554 & {\bf 0.3302} & 0.3389 & {\it 0.3307} \\ \hline
\end{tabular}
}
\caption{Medians and MADs of the RMSE for a rank-two model in $\mathbb{R}^4$ of $50$ points with  Cauchy noise and $20\%$ of missing values according to MCAR, over $1000$ repetitions.}
\label{tab:lowrank50}
\end{center}
\end{table}

\subsubsection{Contamination in higher dimensions}\label{ssec:large}

To check the resistance to outliers in higher dimensions, we consider a simulation setting similar to that of Section~\ref{ssec:StudentOutl}, in dimension $6$, with a normal multivariate distribution with $\bmm_4=(0,\ldots,0)^\top$ and a Toeplitz covariance matrix $\bmS_4$ (having $\sigma_{i,j}=2^{-|i-j|}$ as entries). The data are contaminated with $15\%$ of outliers and have $15\%$ of MCAR values on non-contaminated data. The Tukey depth is approximated using $1000$ random directions. Figure~\ref{fig:mar} (right) shows that the Tukey depth imputation has high predictive quality, comparable to that of the random forest imputation even with only $1000$ random directions.

\subsubsection{Skewed distributions and distributions with non-convex support}\label{ssec:skewednc}

\begin{figure}[!h]
  \begin{center}
    \includegraphics[keepaspectratio=true,height=0.325\textwidth,trim=10mm 9mm 0mm 16mm,clip=true]{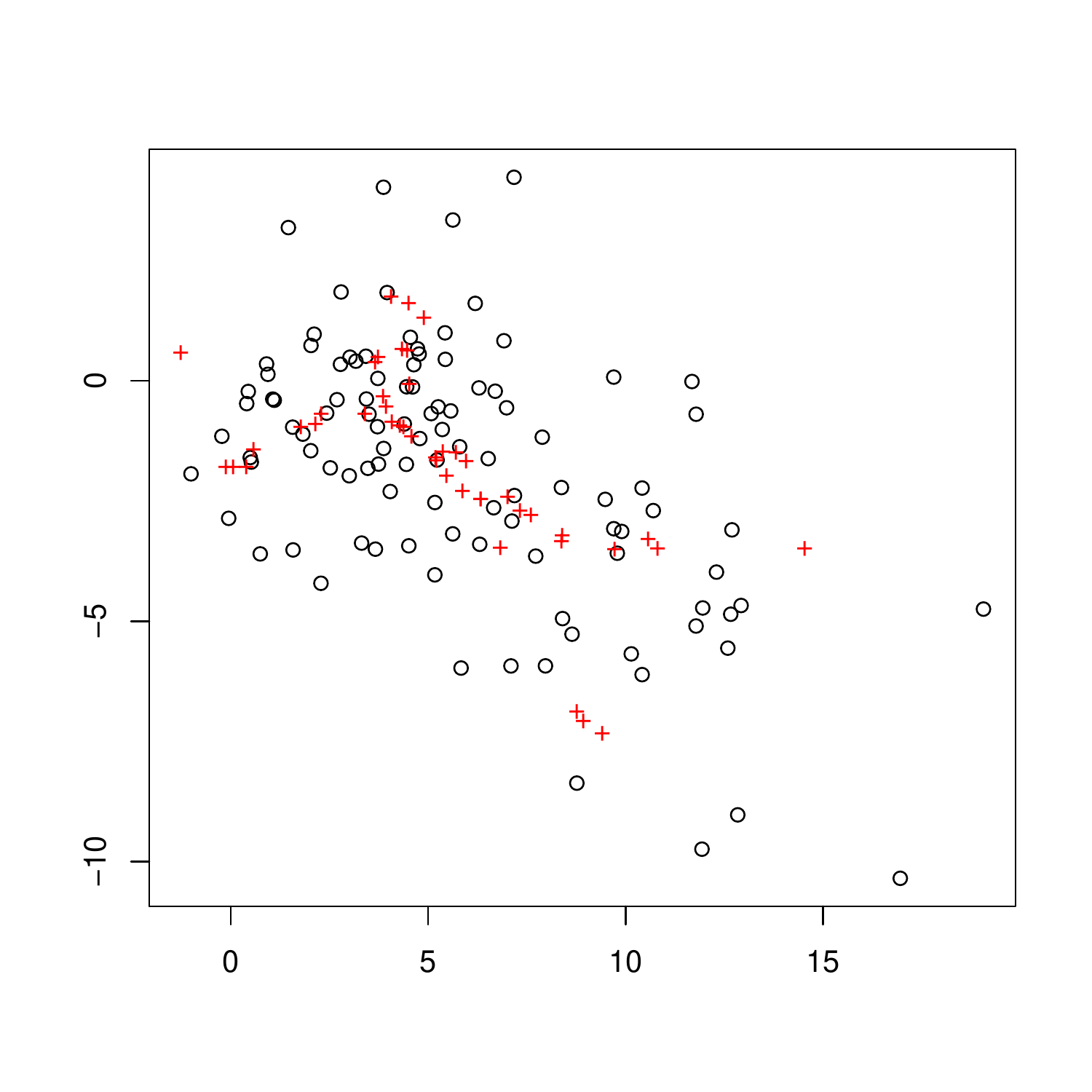}\includegraphics[keepaspectratio=true,height=0.325\textwidth,trim=0mm 5mm 0mm 15mm,clip=true]{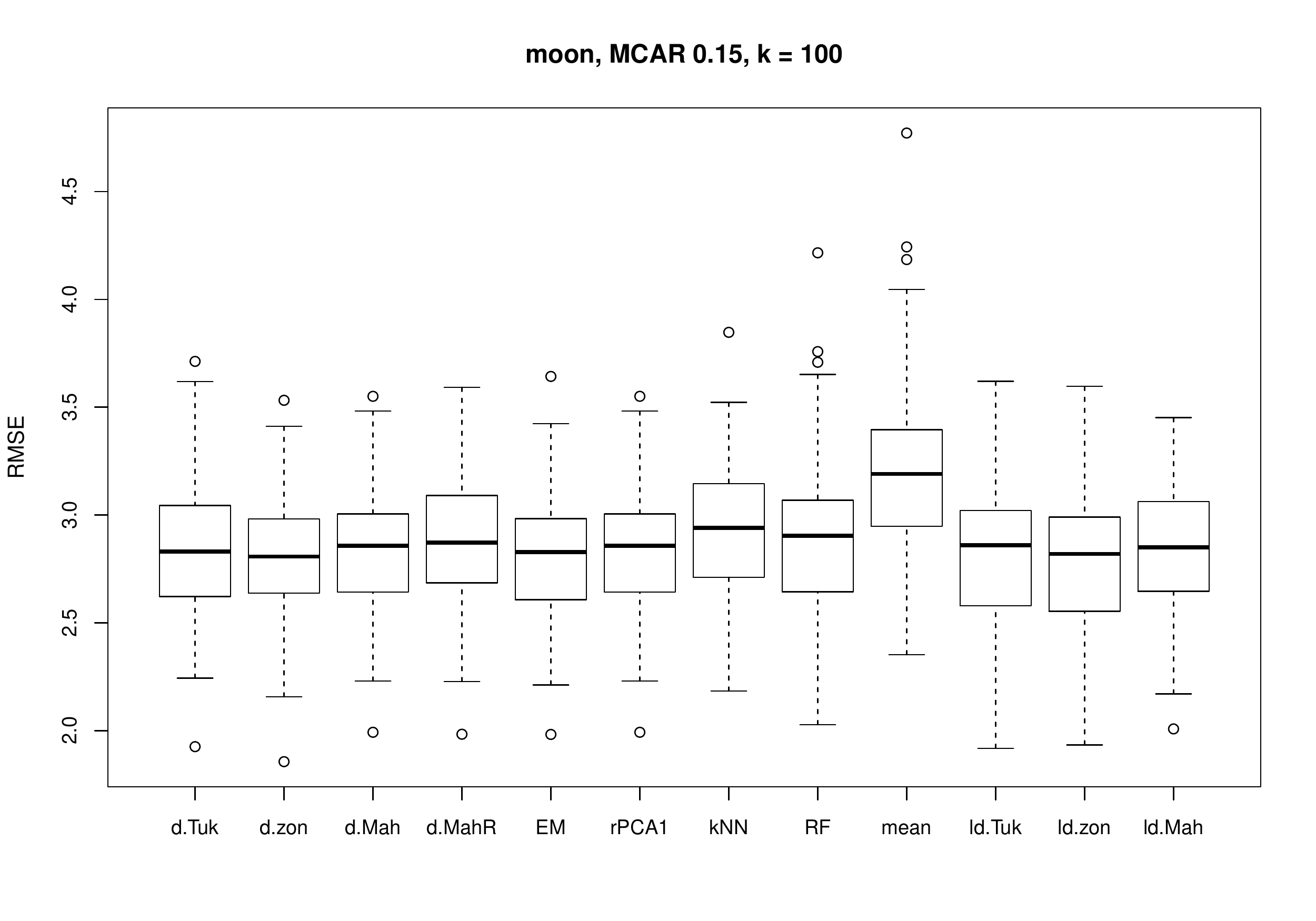}
  \caption{Left: An example of Tukey depth imputation (pluses). Right: boxplots of RMSEs of the prediction for 150 points drawn from a skewed distribution with $15\%$ MCAR, over 100 repetitions; ld.* stands for the local depth with $\beta=0.8$.}
  \label{fig:skewed}
  \end{center}
\end{figure}

First, let us consider only a slight deviation from ellipticity. We simulate $150$ points from a skewed normal distribution~\citep{AzzaliniC99}, insert $15\%$ MCAR values, and impute them with global (Tukey, zonoid and Mahalanobis) depths and their local versions (see Section~\ref{ssec:depthLocal}). This is shown in Figure~\ref{fig:skewed}. In this setting, both global and local imputation perform similarly.

\begin{figure}[!h]
  \begin{center}
    \includegraphics[keepaspectratio=true,height=0.325\textwidth,trim=10mm 9mm 0mm 16mm,clip=true]{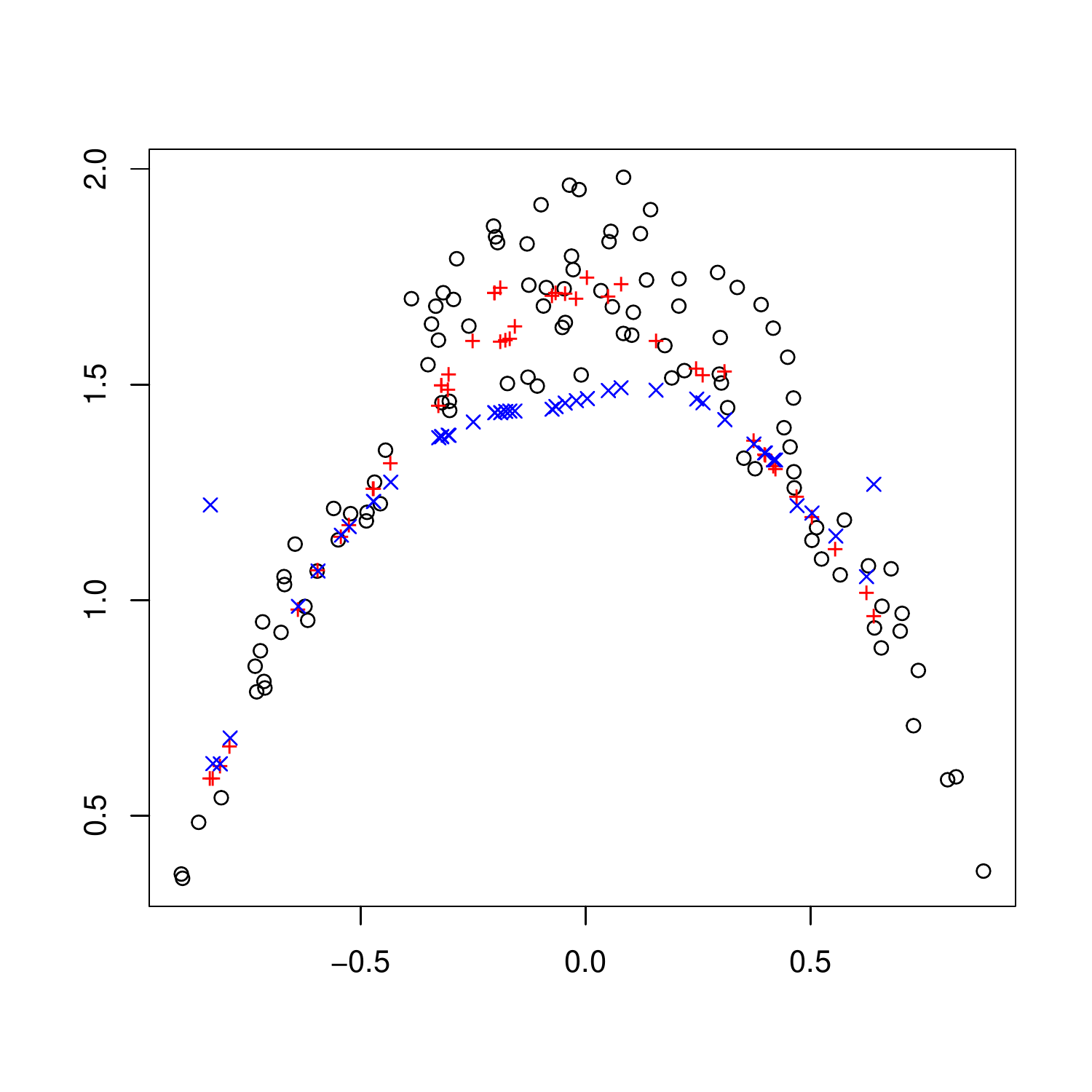}\includegraphics[keepaspectratio=true,height=0.325\textwidth,trim=0mm 5mm 0mm 15mm,clip=true]{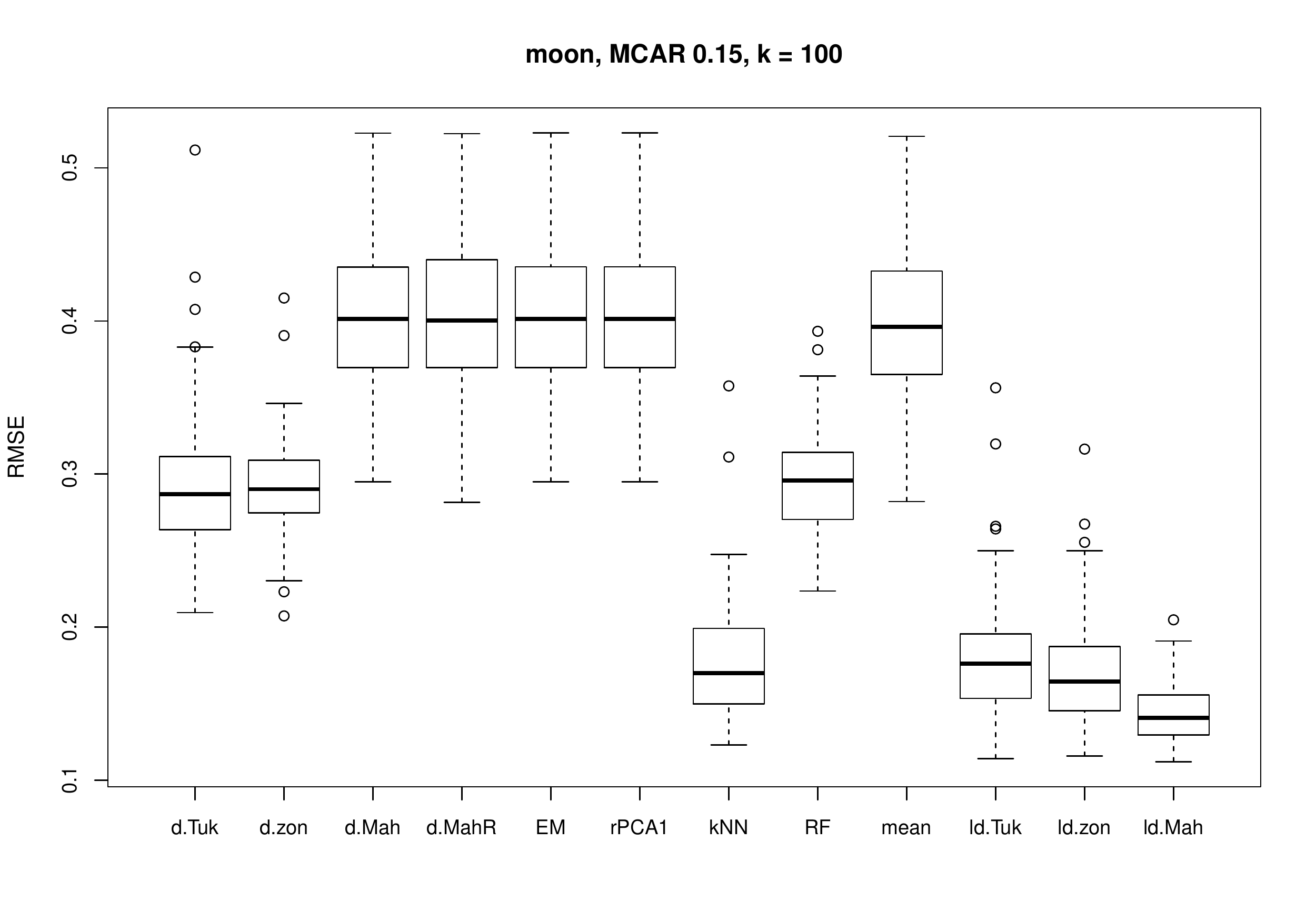}
  \caption{Left: Comparison of global (crosses) and local (pluses) Tukey depth imputation. Right: boxplots of RMSEs of predictions for 150 points drawn from the moon-shaped distribution with $15\%$ MCAR values in the second coordinate, over 100 repetitions; ld.* stands for the local depth with $\beta=0.2$.}
  \label{fig:moon}
  \end{center}
\end{figure}

Further, let us consider an extreme departure from elliplicity with the moon-shaped example from \cite{PaindaveineVB13}. We generate $150$ bivariate observations from $(X_1,X_2)^\top$ with $X_1\sim U(-1,1)$ and $X_2|X_1=x_1\sim U\bigl(1.5(1 - x_1^2),2(1 - x_1^2)\bigr)$, and introduce $15\%$ of MCAR values on $X_2$, see Figure~\ref{fig:moon} (left).
Figure~\ref{fig:moon} (right) shows boxplots of the RMSE for single imputation using local Tukey, zonoid and Mahalanobis depths. If the depth and value of $\beta$ are properly chosen (this can be achieved by cross-validation), the local-depth imputation considerably outperforms the classical methods as well as the global depth.

\subsection{Real data}\label{ssec:expRealdata}

\begin{figure}[!h]
\begin{center}
\includegraphics[keepaspectratio=true,scale=0.335,trim=0mm 15mm 0mm 15mm,clip=true]{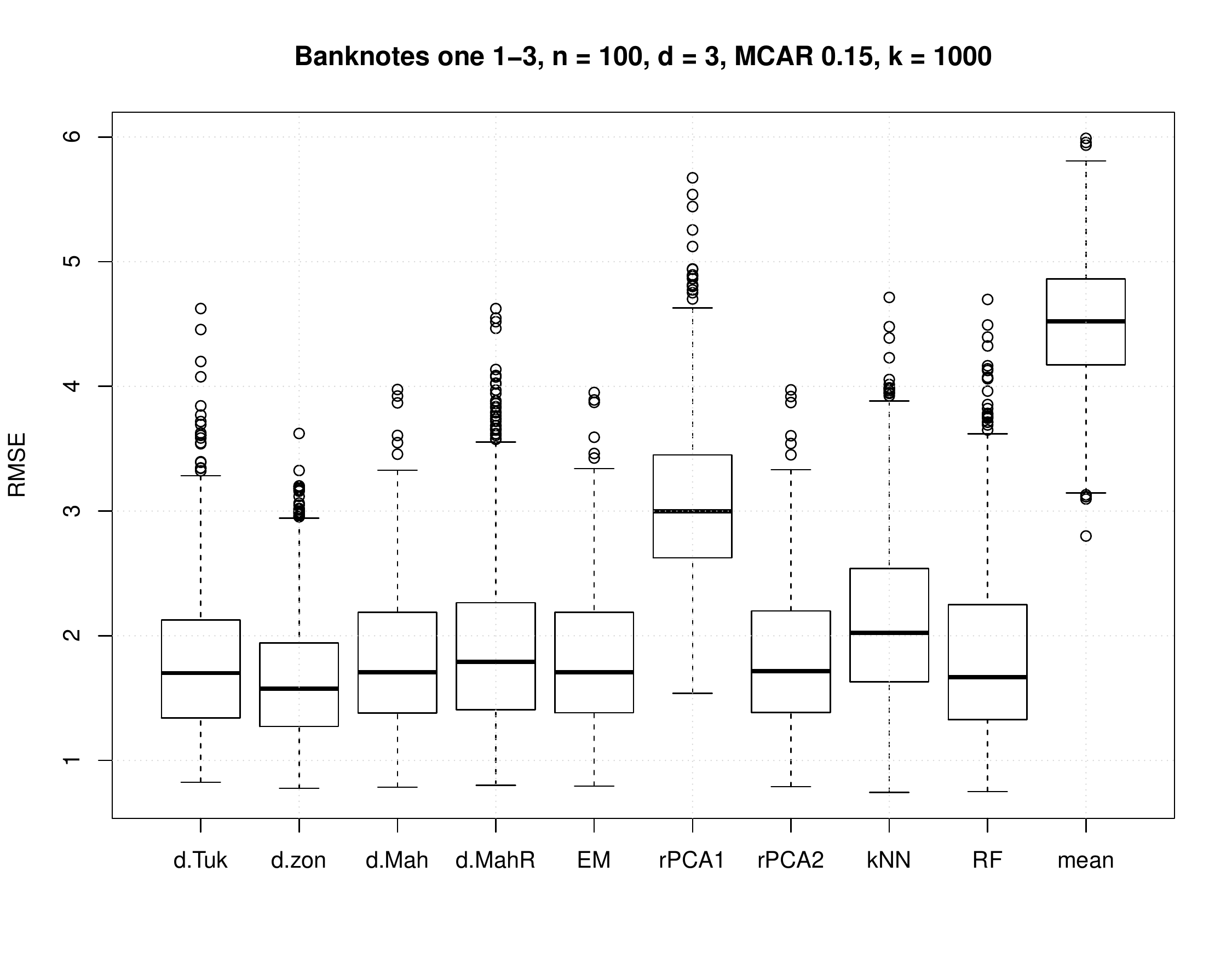}\includegraphics[keepaspectratio=true,scale=0.335,trim=10mm 15mm 0mm 15mm,clip=true]{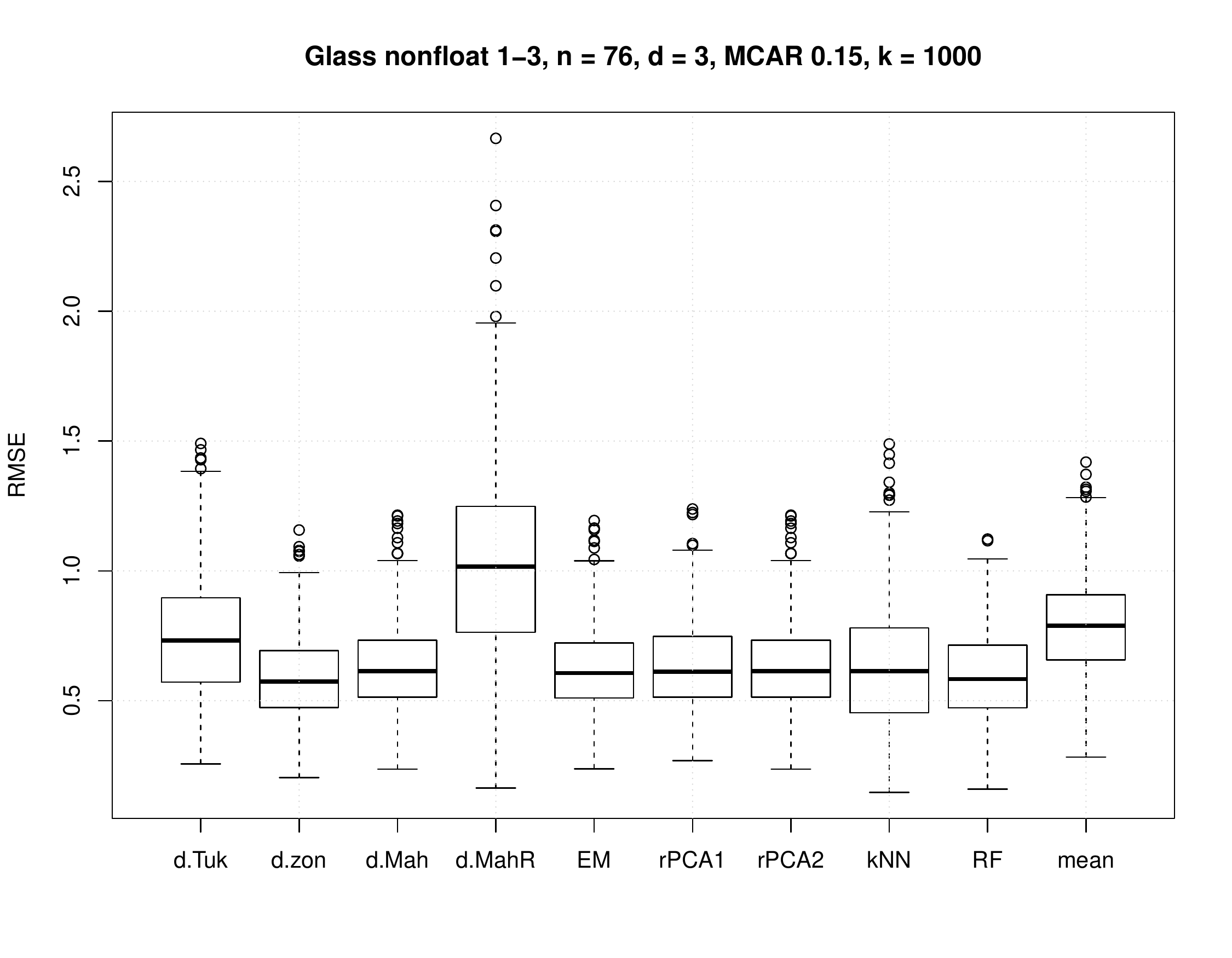}

\includegraphics[keepaspectratio=true,scale=0.335,trim=0mm 15mm 0mm 15mm,clip=true]{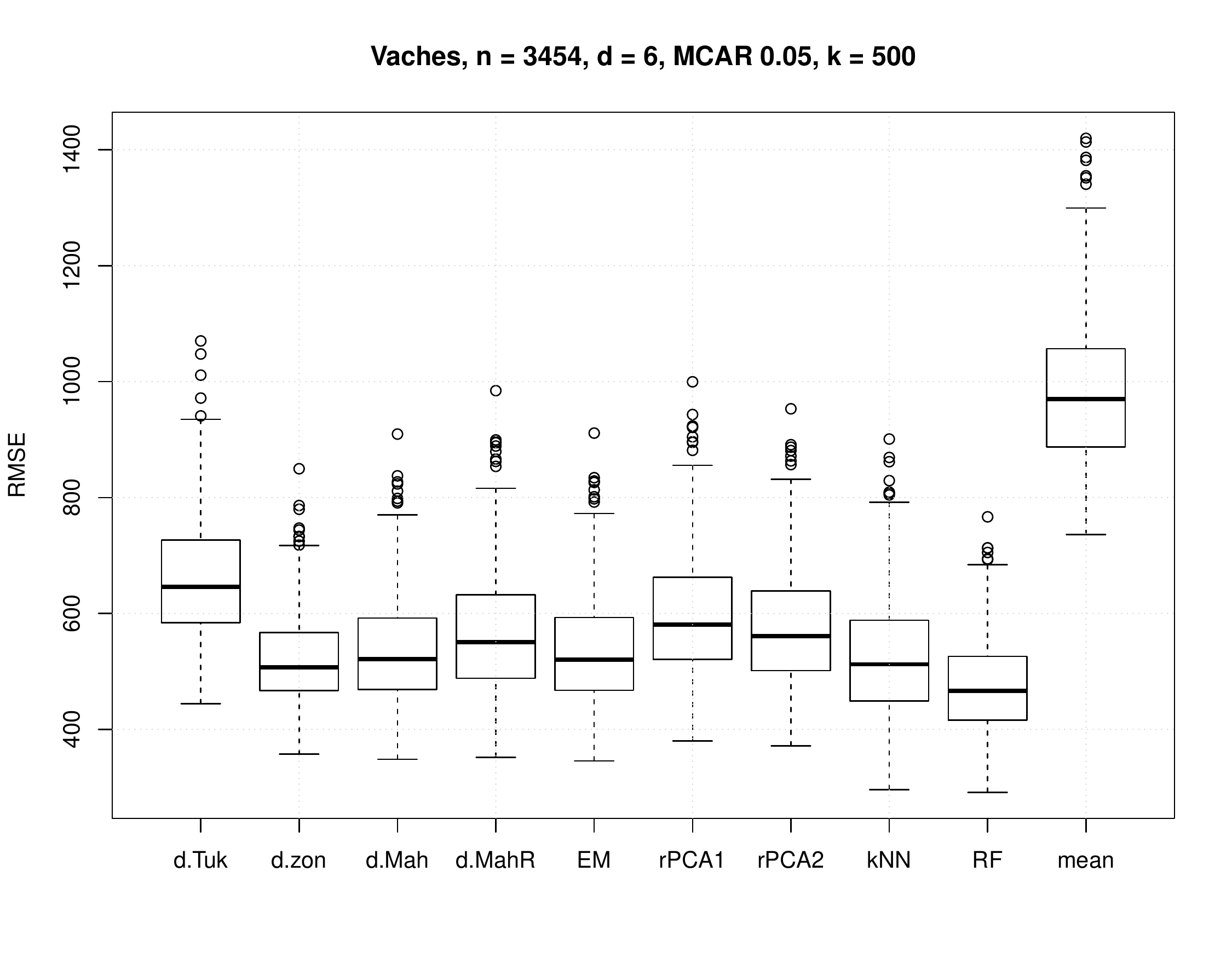}\includegraphics[keepaspectratio=true,scale=0.335,trim=10mm 15mm 0mm 15mm,clip=true]{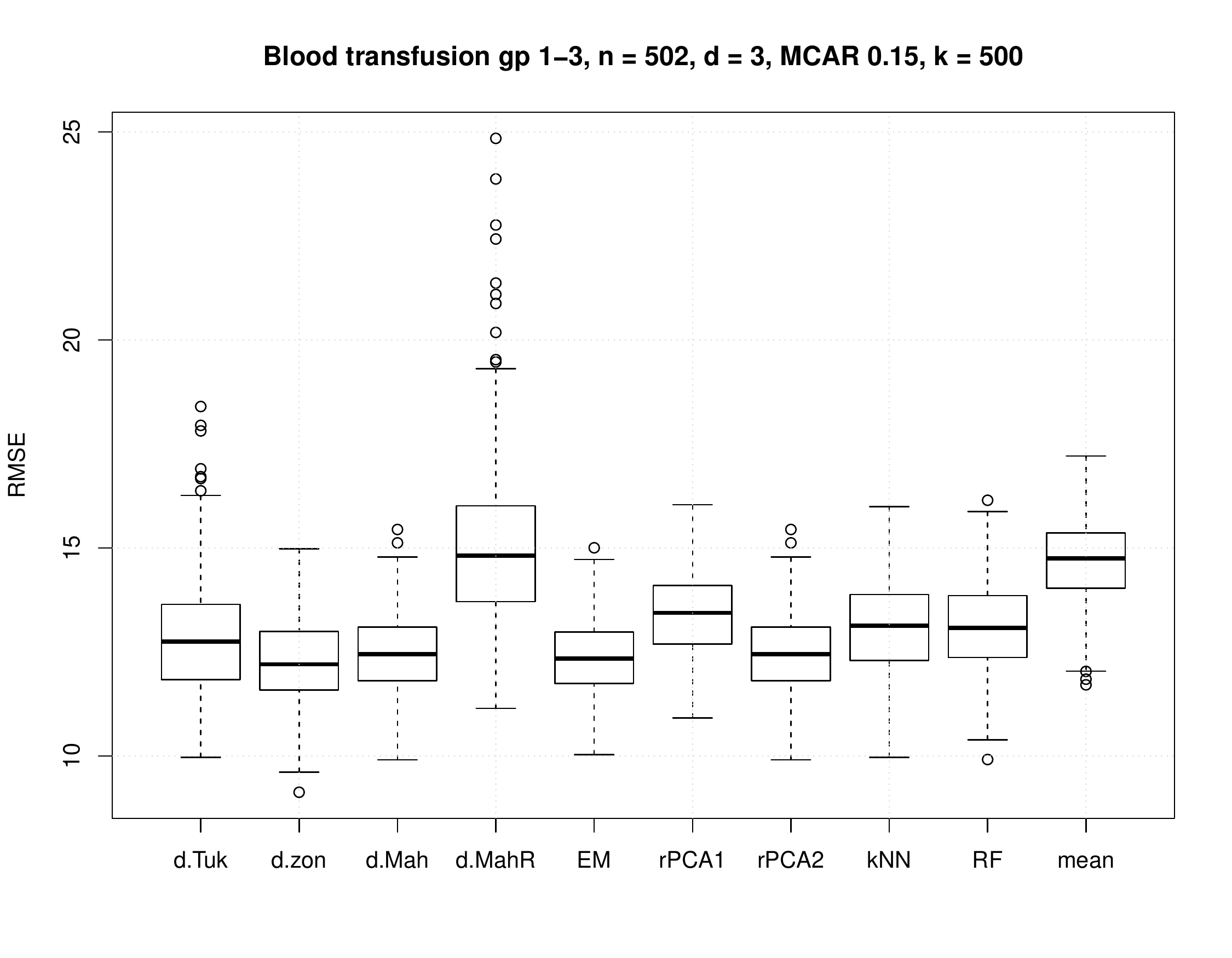}
\caption{RMSEs for the Banknotes (top, left), Glass (top, right), Cows (bottom, left), and Blood Transfusion (bottom, right) data sets with $15\%$ ($5\%$ for Cows) of MCAR values over $500$ repetitions.}
\label{fig:realdata}
\end{center}
\end{figure}

We validate the proposed methodology on three real data sets taken from the UCI Machine Learning Repository~\citep{DuaKT17} and on the  Cows data set.
We thus consider Banknotes ($n=100$, $d=3$), Glass ($n=76$, $d=3$), Blood Transfusion \citep[$n=502$, $d=3$,][]{YehYT08}, and Cows ($n=3454$, $d=6$).
For  details on the experimental design, see the implementation.
Figure~\ref{fig:realdata} shows boxplots of the RMSEs for the ten  imputation methods considered. The zonoid depth is stable across data sets and provides the best results.

Observations in Banknotes are clustered in two groups, which explains the poor performance of the mean and one-dimensional regularized PCA imputation. The zonoid depth searches for a compromise between local and global features and performs the best. The Tukey depth captures the data geometry, but under-exploits information on points' location. Methods imputing by conditional mean (Mahalanobis depth, EM-based, and regularized PCA imputation) perform similarly and reasonably well while imputing in two-dimensional affine subspaces.
The Glass data is challenging as it highly deviates from ellipticity, and part of the data lie sparsely in part of the space, but do not seem to be outlying. Thus, the mean, and robust Mahalanobis and Tukey depth imputation perform poorly. Accounting for local geometry, random forest and zonoid depth perform slightly better.
For the Cows data, which is larger-dimensional, the best results are obtained with random forest, but followed closely by zonoid depth imputation which reflects the data structure. The Tukey depth with $1000$ directions struggles, while the satisfactory results of EM suggest that the data are close to elliptical.
The Blood Transfusion data visually resemble a tetrahedron dispersed from one of its vertices. Thus, mean imputation can be substantially improved. Nonparametric methods and rank one regularized PCA perform poorly because they disregard dependency between dimensions. Better imputation is delivered by those capturing correlation: the depth- and EM-based methods.






Table \ref{tab:time} shows the time taken by different imputation methods. Zonoid imputation is very fast, and the approximation scheme by~\cite{Dyckerhoff04} allows for a scalable application of the Tukey depth.

\begin{table}
	\begin{center}
		  \begin{tabular}{l|cccc}
			Data set & $D^{Tuk}$ & $D^{zon}$ & $D^{Mah}$ & $D^{Mah}_{MCD}$ \\ \hline
			{Higher dimension  (Section~\ref{ssec:large}) ($n=1000$, $d = 6$)} & 3230$^*$ & 1210 & 0.102 & 1.160 \\
			{Banknotes ($n=100$, $d=3$)} & 81.2 & 0.376 & 0.010 & 0.126 \\
			{Glass ($n=76$, $d=3$)} & 12.6 & 0.143 & 0.008 & 0.085 \\
			{Cows ($n=3454$, $d=6$)} & 4490$^*$ & 14300 & 0.212 & 2.37 \\
			{Blood transfusion ($n = 502$, $d = 3$)} & 26400 & 51.3 & 0.46 & 0.775 \\
  		\end{tabular}	
	\end{center}
\caption{Median (in seconds, over $35$ runs) execution time for  depth-based imputation. $^*$~indicates approximate Tukey depth with $1000$ random directions.}
\label{tab:time}
  \end{table}

\section{Multiple imputation for the elliptical family}\label{sec:multiple}


When the objective is to predict missing entries  as well as possible, single imputation is well suited. When analyzing complete data, it is important to go further, so as to better reflect the uncertainty in predicting missing values. This can be done with multiple imputation (MI) \citep{LittleR02} where several plausible values are generated for each missing entry, leading to several imputed data sets. MI then applies a statistical method to each imputed data set, and aggregates the results for inference.
Under the Gaussian assumption, the generation of several imputed data sets is achieved by drawing missing values from the Gaussian conditional distribution of the missing entries, e.g., imputing $\bmx_{miss}$ by draws from $\mathcal{N}(\bmm,\bmS)$ conditional on $\bmx_{obs}$, with the mean and covariance matrix estimated by EM. This method is called stochastic EM.
The objective is to impute close to the underlying distribution. 
However, this is not enough to perform \textit{proper} \citep{LittleR02} multiple imputation, since  uncertainty in the imputation model's  parameters must also be reflected. This is usually obtained either using a bootstrap or Bayesian approach, see e.g., \cite{Schafer97,Efron94, VanBuuren12} for more details.

The  generic framework of depth-based single imputation developed above allows for multiple imputation to be extended to the more general elliptical framework. We first show how to reflect the  uncertainty due to the distribution (Section~\ref{ssec:multImproper}), then apply bootstrap to reflect  model uncertainty, and state the complete algorithm (Section~\ref{ssec:multProper}).


\subsection{Stochastic single depth-based imputation}\label{ssec:multImproper}







\begin{figure}[!h]\centering
    \includegraphics[keepaspectratio=true,scale=0.635]{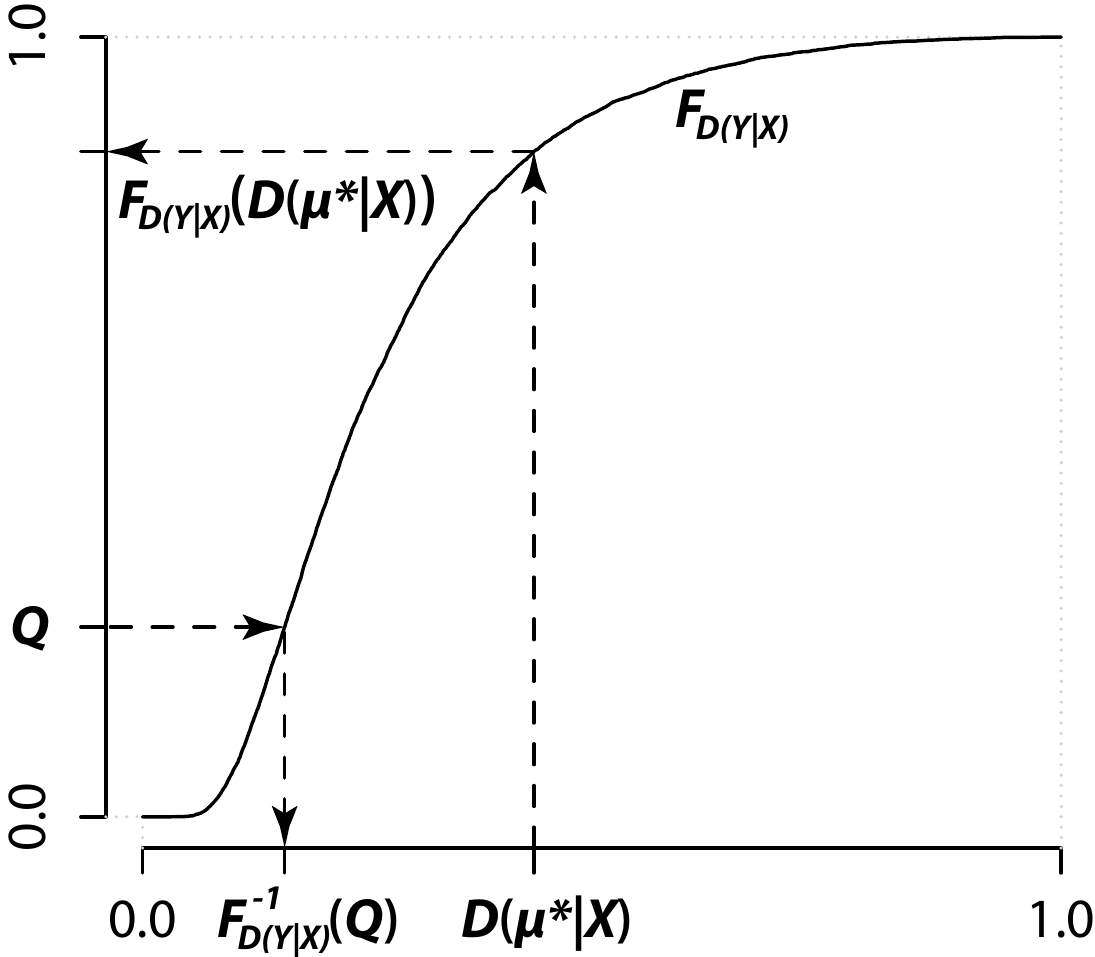}$\quad{}\quad{}$\includegraphics[keepaspectratio=true,scale=.725]{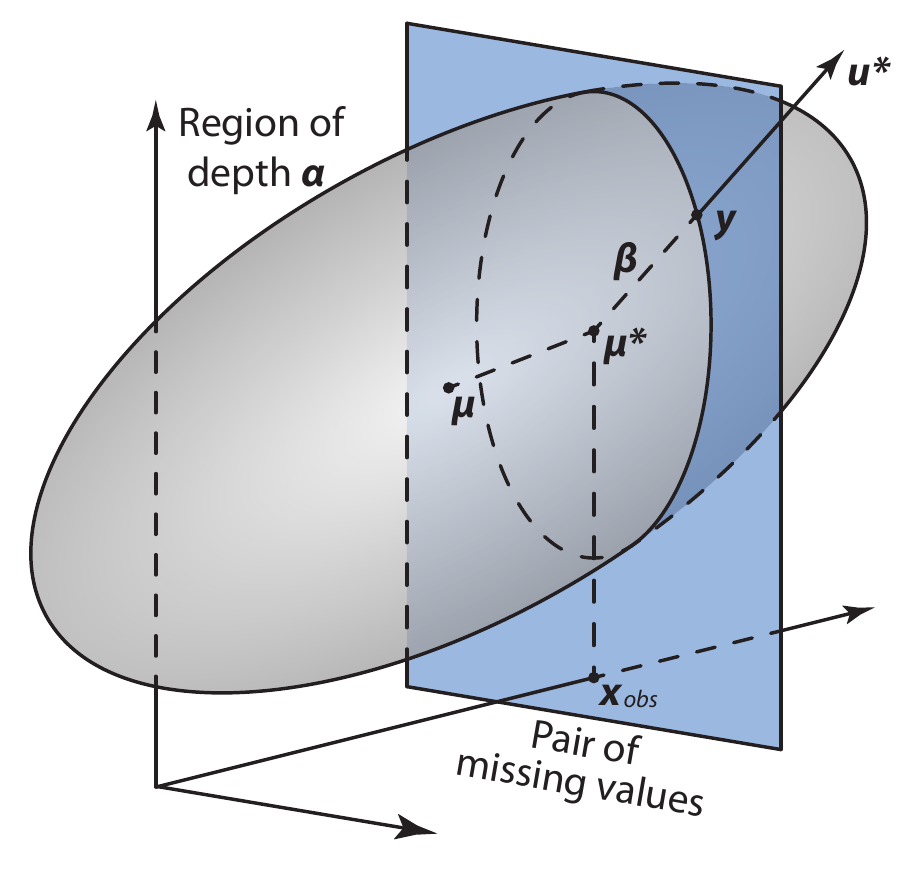}
	\caption{Illustration of an application of (\ref{eqn:depthcdf}) to impute by drawing from the conditional distribution of an elliptical distribution. Drawing the depth $D=F^{-1}_{D(Y|X)}(Q)$ via the depth cumulative distribution function $F_{D(Y|X)}$ (left) and locating the corresponding imputed point $\bmy$ (right).}
	\label{fig:ellipse}
\end{figure}

The extension of  stochastic EM to the elliptically symmetric distribution consists in drawing from a conditional distribution that is also elliptical.
For this we design a  Monte Carlo Markov chain (MCMC), see Figure~\ref{fig:ellipse} for an illustration of a single iteration. First, starting with a point with missing values and observed values $\bmx_{obs}$, we impute it with $\bmm^{*}$ by maximizing its depth \eqref{equ:impAve}, see Figure~\ref{fig:ellipse} (right). Then, for each $\bmy$ with $\bmy_{obs}=\bmx_{obs}$ it holds that $D(\bmy|\bmX) \le D(\bmm^{*}|\bmX)$. The cumulative distribution function (with the normalization constant omitted as it is used, exceptionally, for drawing random variables) of the depth of the random vector $Y$ corresponding to $\bmy$ can be written as 
\begin{equation}
\begin{aligned}
    F_{D(Y|X)}(y) & \, = \, \int_0^y f_{D(X|X)}(z)\frac{\Bigl(\sqrt{d_{M}^2(z) - d_{M}^2\bigl(D(\bmm^{*}|X)\bigr)}\Bigr)^{|miss(\bmx)| - 1}}{d_{M}^{d - 1}(z)} \times\\
    & \, \times \, \frac{d_{M}(z)}{\sqrt{d_{M}^2(z) - d_{M}^2\bigl(D(\bmm^{*}|X)\bigr)}}dz,
\end{aligned}
\label{eqn:depthcdf}
\end{equation}
where $f_{D(X|X)}$ denotes the density of the depth for a random vector $X$ w.r.t. itself, and $d_{M}(z)$ is the Mahalanobis distance to the center as a function of depth (see the supplementary materials for the derivation of this). For the specific case of the Mahalanobis depth, $d_{M}(x)=\sqrt{1/x - 1}$. Then, we draw a quantile $Q$ uniformly on $[0,F_{D(Y|X)}\bigl(D(\bmm^{*}|\bmX)\bigr)]$ that gives the value of the depth of $\bmy$ as $\alpha=F^{-1}_{D(Y|X)}(Q)$, see Figure~\ref{fig:ellipse} (left). $\alpha$ defines a depth contour, which is depicted as an ellipsoid in Figure~\ref{fig:ellipse} (right). Finally, we draw $\bmy$ uniformly in the intersection of this contour with the hyperplane of missing coordinates: $\bmy\in \partial D_\alpha(\bmX)\cap\{\bmz\in\mathbb{R}^d\,|\,\bmz_{obs(\bmx)}=\bmx_{obs}\}$. This is done by drawing $\bmu$ uniformly on $\mathcal{S}^{|miss(\bmx)| - 1}$ and transforming it using a conditional scatter matrix, obtaining $\bmu^*\in\mathbb{R}^d$, where  $\bmu^*_{miss(\bmx)}=\boldsymbol{\Lambda}\bmu$ (with $\boldsymbol{\Lambda}(\boldsymbol{\Lambda})^\top=\bmS_{miss(\bmx),miss(\bmx)} - \bmS_{miss(\bmx),obs(\bmx)}\bmS_{obs(\bmx),obs(\bmx)}^{-1}\bmS_{obs(\bmx),miss(\bmx)}$) and $\bmu^*_{obs(\bmx)}=\boldsymbol{0}$. Such a $\bmu^*$ is uniformly distributed on the conditional depth contour. Then $\bmx$ is imputed as $\bmy=\bmm^{*} + \beta\bmu^*$, where $\beta$ is a scalar obtained as the positive solution of $\bmm^{*} + \beta\bmu^*\in\partial D_\alpha(\bmX)$ (e.g., the quadratic equation $(\bmm^{*} + \beta\bmu^* - \bmm)^\top\bmS^{-1}(\bmm^{*} + \beta\bmu^* - \bmm)=d^2_{M}(\alpha)$ in the case of the Mahalanobis depth), see Figure~\ref{fig:ellipse} (right).

\subsection{Depth-based multiple imputation}\label{ssec:multProper}



We use a bootstrap approach to reflect uncertainty due to the estimation of the underlying semiparametric model. The depth-based procedure for multiple imputation (called DMI), detailed in  Algorithm~\ref{alg:multiple} consists of the following steps: first, a sequence of indices $\bmb=(b_1,\ldots,b_n)$ is drawn from $b_i\sim U(1,\ldots,n)$ for $i=1,\ldots,n$, and this sequence is used to obtain new incomplete data set $\bmX_{\bmb,\cdot}=(\bmx_{b_1},\ldots,\bmx_{b_n})$. Then, on each incomplete data set, the stochastic single depth imputation method described in Section~\ref{ssec:multImproper}  is applied.


\noindent\hspace*{0\parindent}%
\begin{minipage}{0.875\textwidth}
\begin{algorithm}[H]
\caption{Depth-based multiple imputation}\label{alg:multiple}
\begin{algorithmic}[1]
\Function{impute.depth.multiple}{$\boldsymbol{X}$, num.burnin, num.sets}
\For{$m=1:num.sets$}
\State $\boldsymbol{Y}^{(m)}\gets$\Call{impute.depth.single}{$\boldsymbol{X}$} \Comment{Start MCMC with a single imputation}
\State $\bmb\gets(b_1,\ldots,b_n)=\bigl(U(1,\ldots,n),\ldots,U(1,\ldots,n)\bigr)$ \Comment{Draw bootstrap sequence}
\For{$k = 1:(num.burnin + 1)$}
    \State $\boldsymbol{\Sigma}\gets\hat{\boldsymbol{\Sigma}}(\boldsymbol{Y}_{\bmb,\cdot}^{(m)})$
    \State Estimate $f_{D(X|X)}$ using $\boldsymbol{Y}^{(m)}$.
    \For{$i = 1:n$}
        \If{$miss(i)\,\ne\,\varnothing$}
        	\State $\bmm^*\gets\Call{impute.depth.single}{\bmx_i,\bmY_{\bmb,\cdot}^{(m)}}$ \Comment{Single-impute point}
            \State $\bmu\gets U(\mathcal{S}^{|miss(i)| - 1})$
            \State $\bmu^*_{miss(i)}\gets \bmu\boldsymbol{\Lambda}$ \Comment{Calculate random direction}
            \State $\bmu^*_{obs(i)}\gets 0$
            \State Calculate $F_{D(Y|X)}$
            \State $Q\gets U\bigl([0,F_{D(Y|X)}\bigl(D(\boldsymbol{\mu}^*|\boldsymbol{Y}_{\bmb,\cdot}^{(m)})\bigr)]\bigr)$ \Comment{Draw depth}
            \State $\alpha\gets F^{-1}_{D(Y|X)}(Q)$
            \State $\beta\gets$ positive solution of $\bmm^{*} + \beta\bmu^*\in\partial D_\alpha(\bmY_{\bmb,\cdot}^{(m)})$.
            \State $\boldsymbol{y}_{i,miss(i)}^{(m)}\gets\boldsymbol{\mu}_{miss(i)}^* + \beta \bmu^*_{miss(i)}$ \Comment{Impute one point}
        \EndIf
    \EndFor
\EndFor
\EndFor
\State\Return $\bigl(\boldsymbol{Y}^{(1)},\ldots,\boldsymbol{Y}^{(num.sets)}\bigr)$
\EndFunction
\end{algorithmic}
\end{algorithm}
\end{minipage}

\indent\\


\subsection{Experiments}

\subsubsection{Stochastic single depth-based imputation preserves quantiles}



We generate 500 points from an elliptical Student-$t$ distribution with $3$ degrees of freedom, with $\bmm_4=(-1, -1, -1, -1)^\top$ and $ \bmS_4=\bigl((0.5, 0.5, 1, 1)^\top,\,(0.5, 1, 1, 1)^\top,\,(1, 1, 4, 4)^\top,\,(1, 1, 4, 10)^\top\bigr)$, adding $30\%$ of MCAR values, and compare the imputation with stochastic EM, stochastic PCA~\citep{JosseH12}, and  stochastic depth imputation (Section~\ref{ssec:multImproper}). On the completed data, we calculate quantiles for each variable and compare them with those obtained for the initial complete data. Table~\ref{tab:qsT3} shows the medians of the results over $2000$ simulations for the first variable; the results are the same for the other variables. The stochastic EM and PCA methods, which generate noise from the normal model, do not lead to accurate quantile estimates. The proposed method gives excellent results with only slight deviations in the tail of the distribution due to difficulties in reflecting the density's shape. Although such an outcome is expected, it considerably broadens the scope of practice in comparison to the deep-rooted Gaussian imputation.

\begin{table}[!h]
\begin{center}
\begin{tabular}{l||llllllll}
\multicolumn{1}{l||}{Quantile:} & \phantom{-}0.5 & \phantom{-}0.75 & \phantom{-}0.85 & 0.9 & 0.95 & 0.975 & 0.99 & 0.995 \\ \hline
complete & -1.0013 & -0.4632 & -0.1231 & 0.1446 & 0.6398 & 1.2017 & 2.0661 & 2.8253 \\
stoch. EM & -1.0008 & -0.4225 & -0.0649 & 0.2114 & 0.6902 & 1.2022 & 1.9782 & 2.6593 \\
stoch. PCA & -0.9999 & -0.4248 & -0.0650 & 0.2121 & 0.7048 & 1.2398 & 2.0417 & 2.7481 \\
depth & -0.9996 & -0.4643 & -0.1232 & 0.1491 & 0.6509 & 1.2142 & 2.0827 & 2.8965 \\
\end{tabular}
\caption{Median (over 2000 repetitions) quantiles of the imputed variable $X_1$ obtained from an elliptical sample of $500$ points drawn from the Student-$t$ distribution with~$3$~d.f. with $30\%$ of MCAR values.}\label{tab:qsT3}
\end{center}
\end{table}

\subsubsection{Inference with missing values}


We explore the performance of DMI for inference with missing values
by estimating coefficients of a regression model. Data are generated according to the following model:
$Y = \boldsymbol{\beta}^\top(1, X^\top)^\top + \epsilon$, with 
$\boldsymbol{\beta}=(0.5,1,3)^\top$ and
$X\sim N\Bigl((1, 1)^\top, \bigl((1, 1)^\top, (1, 4)^\top\bigr)\Bigr)$, then 
$30\%$ of MCAR values are introduced.
We employ DMI
and perform multiple imputation using the \texttt{R}-packages \texttt{Amelia} and \texttt{mice} under their default settings, generating $20$ imputed data sets. For each, we run the regression model to estimate the parameters and their variance, and  combine the results according to Rubin's rules~\citep{LittleR02}. Here competitors are in a favourable setting as they are based on Gaussian distribution assumptions. 
We indicate the medians, the coverage of the $95\%$ confidence interval,  and the width of this interval, for the estimates of $\boldsymbol{\beta}$ in Table~\ref{tab:mulReg2} with a sample size of 500, over 2000 simulations.
In addition to the missing data, one difficulty comes from the high correlation ($\approx $0.988) between two of the variables. 
\begin{table}[!h]
\begin{center}
\begin{tabular}{l||lll||lll||lll}
 & \multicolumn{3}{c||}{$\beta_0$} & \multicolumn{3}{c||}{$\beta_1$} & \multicolumn{3}{c}{$\beta_2$} \\ \hline
 & med & cov & width & med & cov & width & med & cov & width \\ \hline
 & \multicolumn{9}{c}{$20$ multiply-imputed data sets} \\ \hline
\texttt{Amelia} & 0.487 & 0.931 & 0.489 & 1.01  & 0.941 & 0.399 & 2.998 & 0.929 & 0.206 \\
\texttt{mice} & 0.519 & 0.984 & 1.6   & 1.081 & 0.98  & 1.807 & 2.881 & 0.982 & 1.502 \\
regPCA & 0.495 & 0.971 & 0.853 & 1.04 & 0.964 & 0.751 & 2.957 & 0.936 & 0.334 \\
DMI & 0.504 & 0.971 & 0.613 & 0.989 & 0.979 & 0.519 & 3.003 & 0.97  & 0.26  \\ \hline
\end{tabular}
\caption{Medians (med), $95\%$ coverage (cov), and width of the confidence intervals (width) for the regression parameters based on $20$ imputed data sets over $2000$ repetitions, for a sample of $500$ observations from a regression model with $30\%$ MCAR values.}\label{tab:mulReg2}
\end{center}
\end{table}

\texttt{Amelia} has minor under-coverage problems, and \texttt{mice} provides biased coefficients and has large over-coverage issues as it is based on regression imputations that are unstable in the presence of high correlation.
DMI, on the other hand, suffers from a slight amount of over-coverage, but in general provides  valid inference. 

\section{Conclusions}\label{sec:conclusions}

The  depth imputation framework we propose here fills the gap between global imputation of regression- and PCA-based methods, and the local imputation of methods such as random forest and $k$NN. It reflects uncertainty in the distribution assumption by imputing data close to the data geometry, is robust in the sense of the distribution and outliers, and still functions with MAR data.
When used with the Mahalanobis depth, using data depth as a concept, the link between iterative regression, regularized PCA, and imputation with values that minimize the determinant of the covariance matrix, was established.
Our empirical study shows the effectiveness of the suggested methodology for various elliptic distributions and  real data.
In addition, the method has been naturally extended to multiple imputation for the elliptical family, broadening the scope of existing tools for multiple imputation.

The methodology is general, i.e., any reasonable notion of data depth can be used, which then determines the  imputation properties. In the empirical study, the zonoid depth behaves well in general, and for real data in particular. However if robustness is an issue, the Tukey depth may be preferable.
The projection depth~\citep{ZuoS00a} is an appropriate choice if only a few points contain missing values in a data set that is substantially outlier-contaminated. This specific case is not included in the article, but imputation based on projection depth is implemented in the associated \texttt{R} package.
To reflect multimodality of the data, the suggested framework has been used with  localized depths, see e.g. \cite{PaindaveineVB13}. 

A serious issue with data depths is their computation. Using approximate versions of data depths (which can also be found in the implementation) is a first step to handling larger data sets.
Our methodology has been implemented as the \texttt{R}-package \texttt{imputeDepth}. Source code of the package and of the experiment-reproducing files can be downloaded from \texttt{https://github.com/julierennes/imputeDepth}. 

\section*{Acknowledgments}

The authors thank co-editor in chief Regina Liu, the associate editor, and  three anonymous reviewers for their insightful remarks. The authors also gratefully thank Davy Paindaveine and Bernard Delyon for fruitful discussions as well as Germain Van Bever for providing the original code for the local depth, which served as a starting point for its implementation in the \texttt{R}-package \texttt{imputeDepth}.

\section*{Supplementary materials}

\begin{description}
	\item[Supplementary materials] These contain additional figures, results on experiments with other percentages of missing values, as well as proofs. \\ (\texttt{NIDD\_Supplement.pdf})
	\item[\texttt{R}-package and codes for reproducing experiments] The archive contains the source code of the accompanying \texttt{R}-package \texttt{imputeDepth} as well as  scripts for reproducing the experiments and figures of the article. The most up-to-date content of this archive can be found online using the link \texttt{https://github.com/julierennes/imputeDepth}. \\ (\texttt{NIDD\_codes.zip}).
\end{description}











\bibliographystyle{agsm}

\clearpage

\setcounter{section}{0}

\begin{center}
    {\LARGE\bf Supplementary Materials to the article ``Nonparametric imputation by data depth''} \\
    \vspace{.5cm}
    
    by Pavlo Mozharovskyi, Julie Josse and Fran\c{c}ois Husson
\end{center}

\section{Additional figures}
\begin{figure}[!th]
\begin{center}
\quad\includegraphics[keepaspectratio=true,scale=0.225]{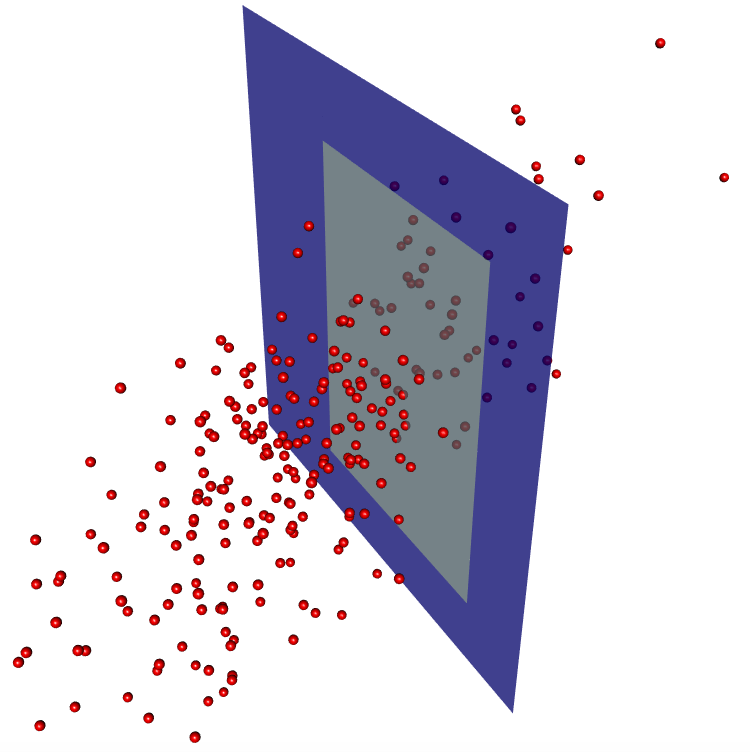}\qquad\,\,\includegraphics[keepaspectratio=true,scale=1.75]{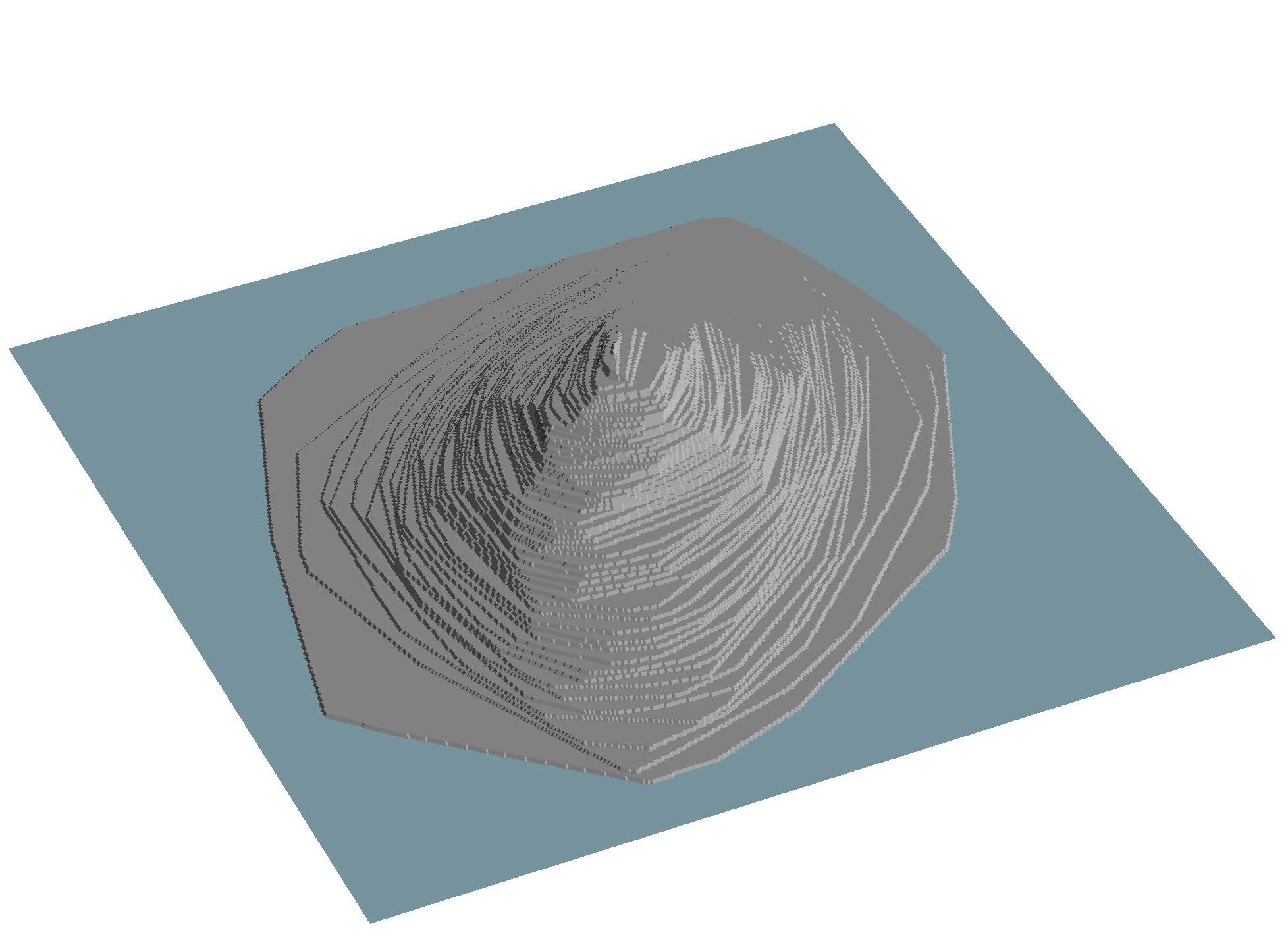}

\indent\\
\includegraphics[keepaspectratio=true,scale=1.75]{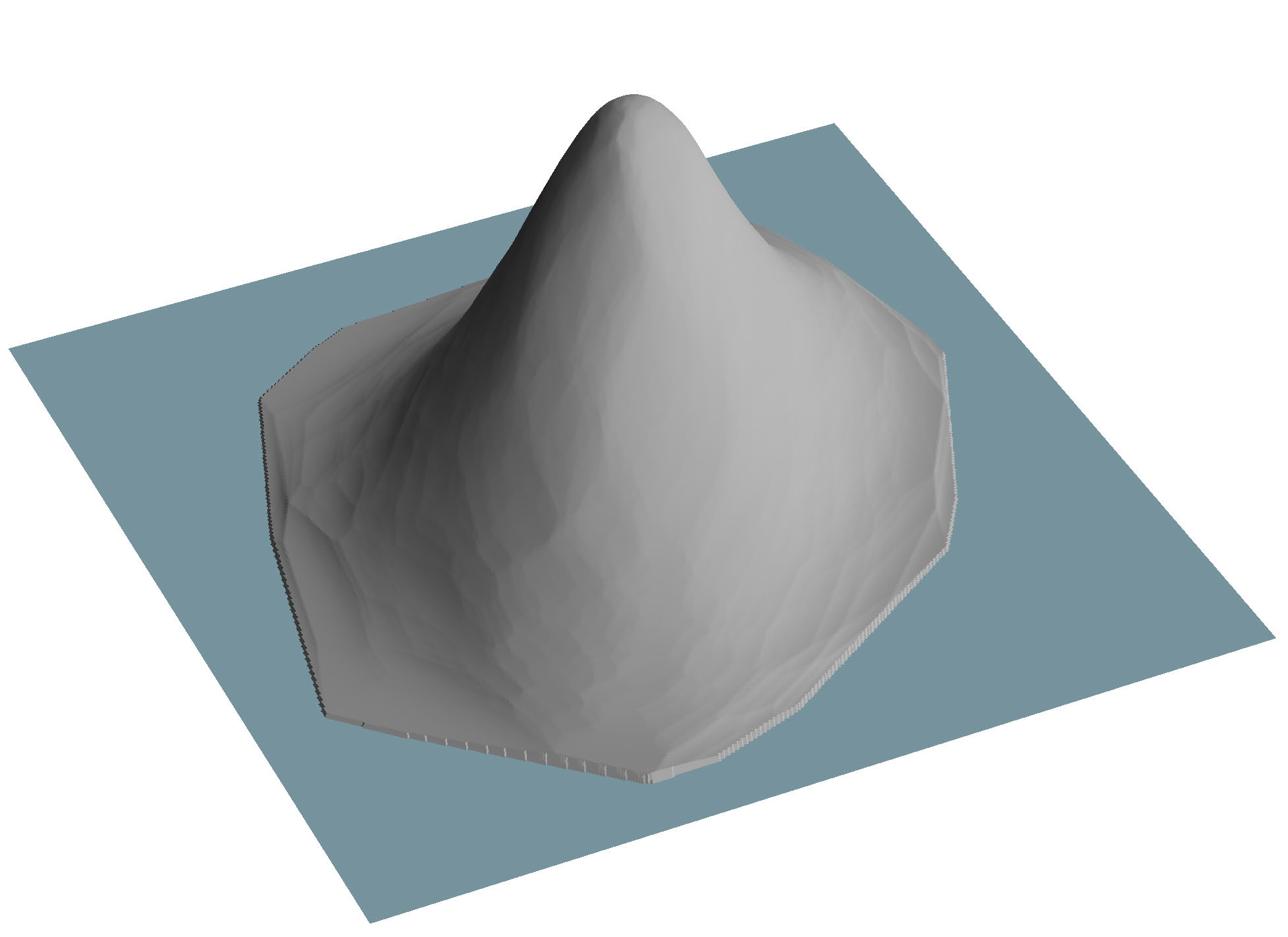}\,\includegraphics[keepaspectratio=true,scale=1.75]{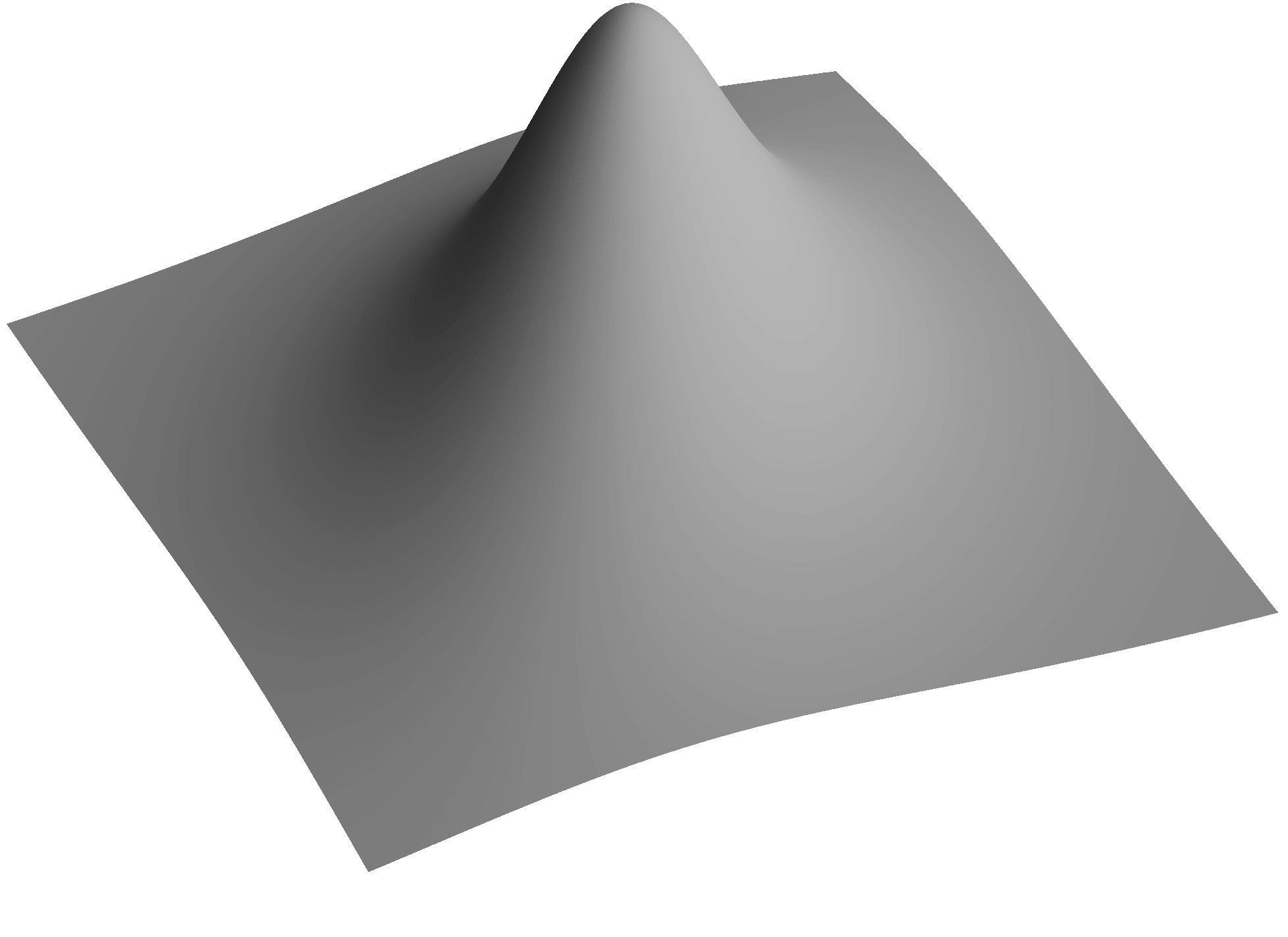}
\caption{A Gaussian sample consisting of $250$ points and a hyperplane of two missing coordinates (top, left), and the function $f(\bmz_{miss})$ to be optimized on each single iteration of Algorithm~1, for the smaller rectangle, for Tukey (top, right), zonoid (bottom, left), and Mahalanobis (bottom, right) depth. For the Tukey depth the maximum is not unique, and forms a polygon.}
\label{fig:optim}
\end{center}
\end{figure}

\begin{figure}[!ht]
\begin{center}
\includegraphics[width=0.435\textwidth, trim=0mm 8mm 0mm 0mm,clip=true]{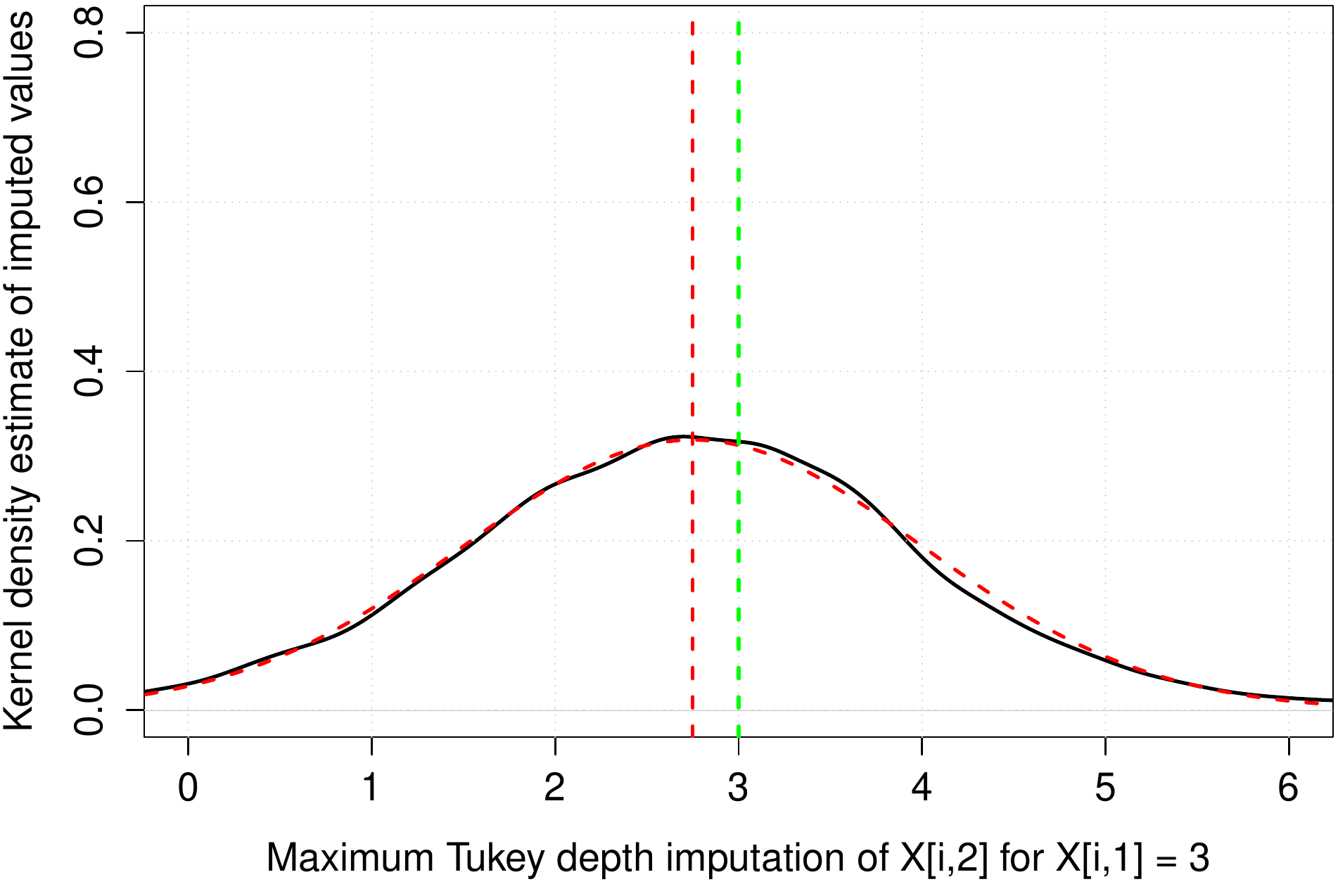}\,\quad\,\includegraphics[width=0.41\textwidth, trim=10mm 8mm 0mm 0mm,clip=true]{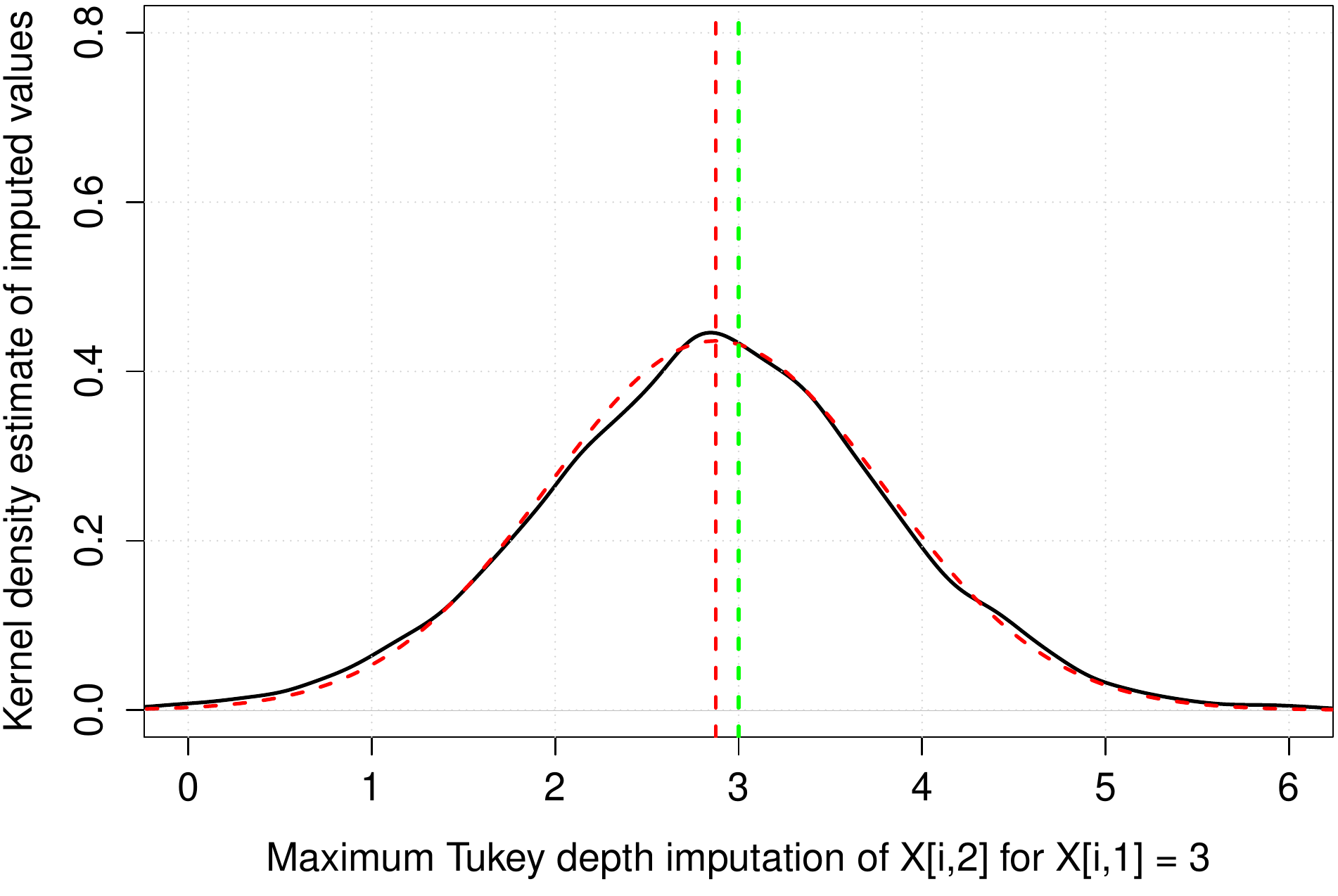}\\
\includegraphics[width=0.435\textwidth, trim=0mm 8mm 0mm 0mm,clip=true]{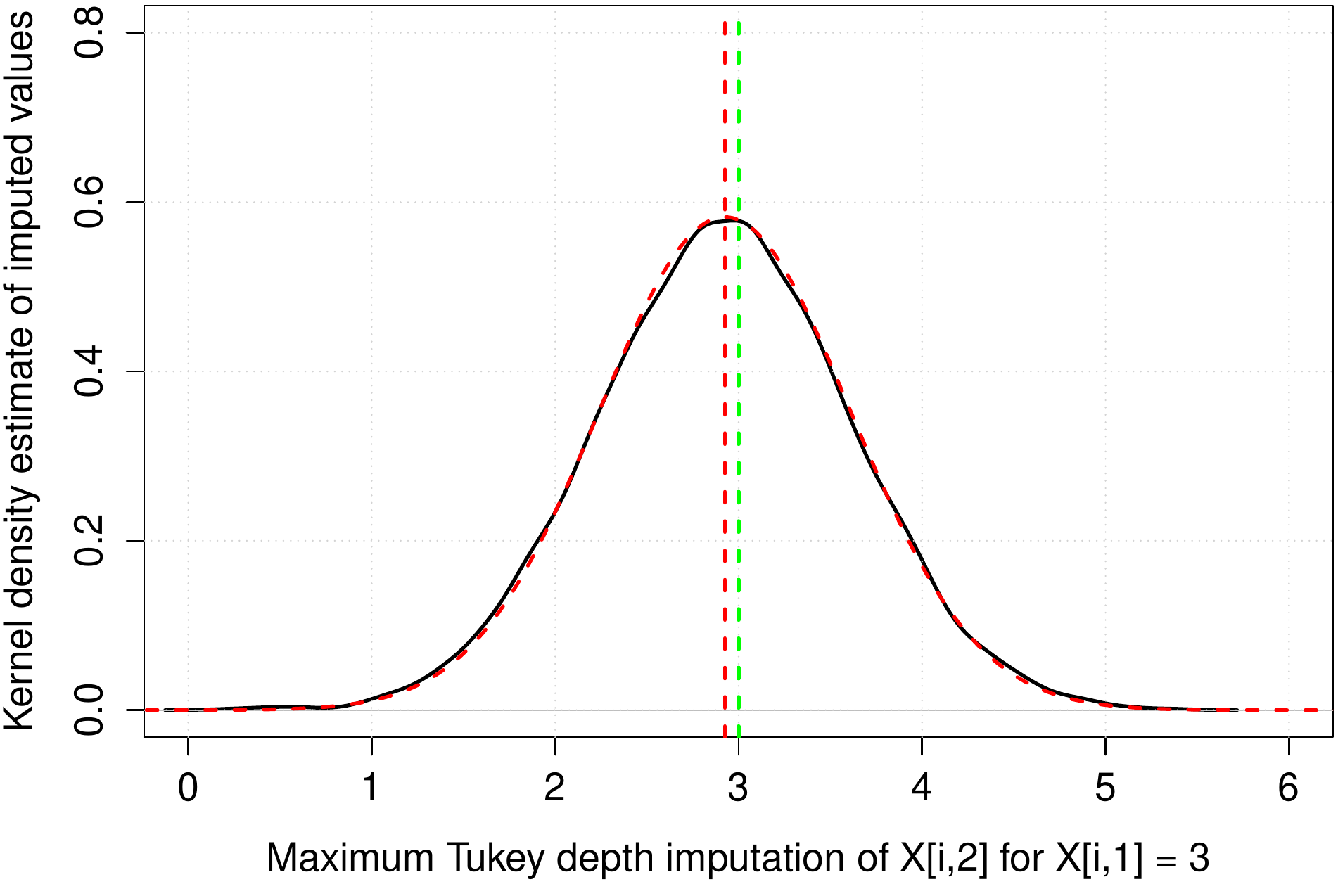}\,\quad\,\includegraphics[width=0.41\textwidth, trim=10mm 8mm 0mm 0mm,clip=true]{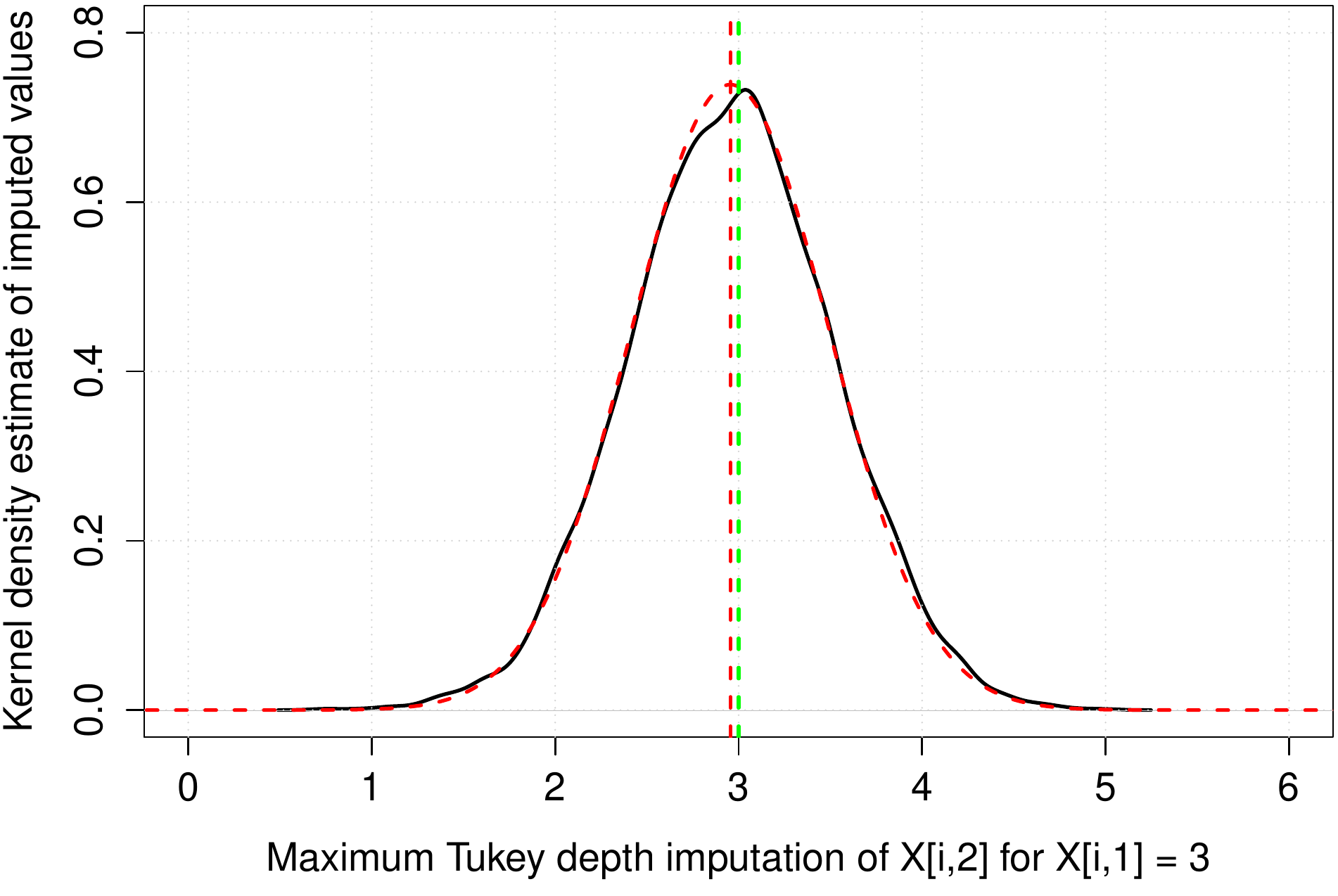}
\end{center}
\caption{Samples of size $100$ (top, left), $200$ (top, right), $500$ (bottom, left), and $1000$ (bottom, right) are drawn from the bivariate Cauchy distribution with the location and scatter parameters $\bmm_1$ and $\bmS_1$ from the introduction. Single point with one missing coordinate is imputed with the Tukey depth. Its kernel density estimate (solid) and the best approximating Gaussian curve (dashed) over $10,000$ repetitions are plotted. The population's conditional center given the observed value equals $3$.}
\label{fig:onerow}
\end{figure}

\clearpage

\begin{figure}[!ht]
\begin{center}
	\includegraphics[keepaspectratio=true,width=0.425\textwidth,trim=10mm 10mm 10mm 10mm,clip=true,page=2]{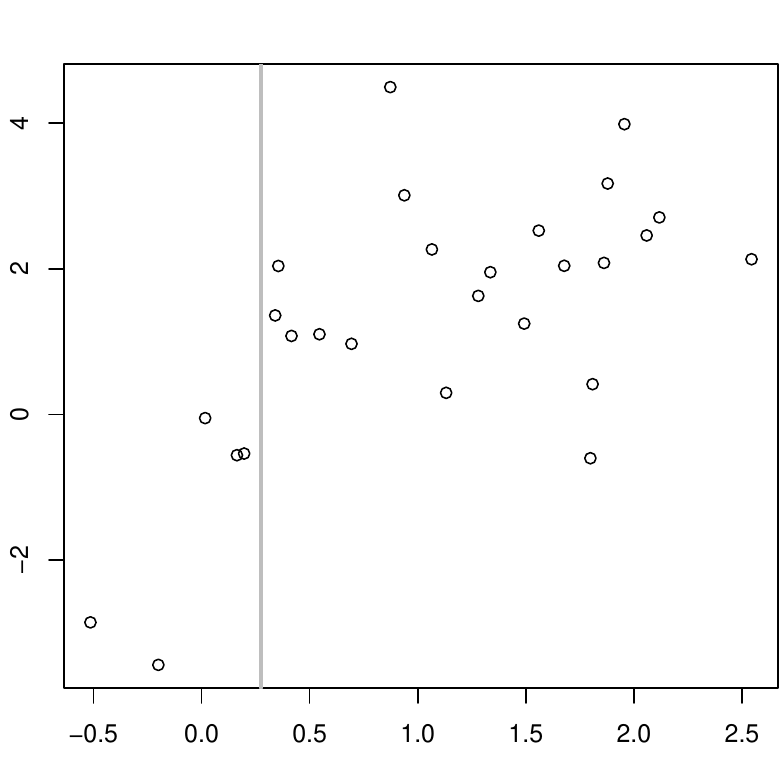}\quad\includegraphics[keepaspectratio=true,width=0.425\textwidth,trim=10mm 10mm 10mm 10mm,clip=true,page=5]{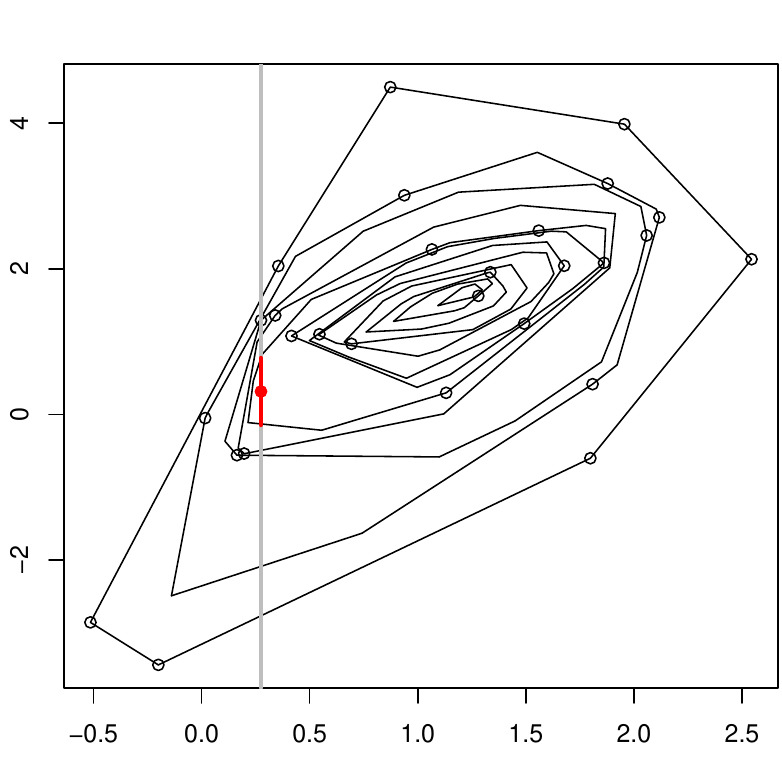}
\end{center}
	\caption{Illustration of imputation with the Tukey depth. When imputing the point with a missing second coordinate (left), the maximum of the constrained Tukey depth is non-unique (the red line segment), and an average over the optimal arguments (the red point) is used in equation~(3) (right).}
	\label{fig:impTukey}
\end{figure}

\section{Simulation results with other percentages of missing values}

When varying the percentage of missing values, the general trend remains unchanged. The small differences seen can be summarized as follows: with a decreasing percentage of missing values, the difference between EM and Mahalanobis depth imputation (and thus also the rank two PCA one) shrinks, and indeed the latter performs comparably to EM for $5\%$ missingness. For the same percentage and the Cauchy distribution, nonparametric methods ($k$NN and random forest)  perform comparably to the Tukey depth due to a sufficient quantity of available observations and an absence of correlation structure (outliers are generated from Cauchy distribution as well).

\begin{table}[!ht]
  \begin{center}
  {\tiny
    \begin{tabular}{|c|c|c|c|c|c|c|c|c|c|c|c|}
    \hline
Distr. & $D^{Tuk}$ & $D^{zon}$ & $D^{Mah}$ & $D^{Mah}_{MCD.75}$ & EM & regPCA1 & regPCA2 & $k$NN & RF & mean & oracle \\
    \hline
$t\,\infty$ & 1.577 & 1.547 & {\it 1.532} & 1.537 & {\bf 1.518} & 1.596 & {\it 1.532} & 1.684 & 1.681 & 2.058 & 1.487 \\
 & (0.2345) & (0.2128) & {\it (0.216)} & (0.2199) & {\bf(0.2129)} & (0.2327) & {\it (0.2159)} & (0.2422) & (0.2445) & (0.2774) & (0.2004) \\ \cline{1-1}
$t\,10$ & 1.748 & 1.718 & {\it 1.693} & 1.709 & {\bf 1.69} & 1.769 & {\it 1.692} & 1.853 & 1.871 & 2.275 & 1.642 \\
 & (0.287) & (0.2838) & {\it (0.2737)} & (0.2827) & {\bf (0.2826)} & (0.3039) & {\it (0.2741)} & (0.3168) & (0.3085) & (0.3757) & (0.2771) \\ \cline{1-1}
$t\,5$ & 1.993 & 1.971 & {\it 1.956} & {\it 1.956} & {\bf 1.933} & 2.017 & {\it 1.956} & 2.125 & 2.134 & 2.565 & 1.874 \\
 & (0.378) & (0.3602) & {\it (0.3799)} & {\it (0.361)} & {\bf (0.361)} & (0.3759) & {\it (0.3796)} & (0.4126) & (0.3976) & (0.4732) & (0.3492) \\ \cline{1-1}
$t\,3$ & 2.417 & 2.434 & 2.39 & {\bf 2.333} & {\it 2.362} & 2.431 & 2.39 & 2.55 & 2.592 & 3.045 & 2.235 \\
 & (0.5996) & (0.6032) & (0.5792) & {\bf (0.5571)} & {\it(0.5808)} & (0.5734) & (0.5793) & (0.6154) & (0.612) & (0.6943) & (0.5319) \\ \cline{1-1}
$t\,2$ & {\it 3.31} & 3.373 & 3.431 & {\bf 3.192} & 3.366 & 3.437 & 3.422 & 3.538 & 3.555 & 4.155 & 2.986 \\
 & {\it (1.191)} & (1.273) & (1.314) & {\bf (1.148)} & (1.249) & (1.343) & (1.289) & (1.33) & (1.321) & (1.466) & (1.063) \\ \cline{1-1}
$t\,1$ & {\bf 13.19} & 15.13 & 15.17 & {\it 13.39} & 14.86 & 14.82 & 15.22 & 14.09 & 13.91 & 16.77 & 11.17 \\
 & {\bf (10.83)} & (12.06) & (11.74) & {\it (10.32)} & (11.57) & (11.64) & (11.94) & (11.28) & (11.06) & (13.22) & (8.901) \\ \hline
    \end{tabular}
  }
  \caption{Median and MAD of the RMSEs of the imputation for a sample of $100$ points drawn from  elliptically symmetric Student-$t$ distributions with  $\bmm_2$ and $\bmS_2$  having $15\%$ of  MCAR values, over 1000 repetitions.}
  \label{tab:StudentT100MCAR15}
  \end{center}
\end{table}

\begin{table}[!ht]
  \begin{center}
  {\tiny
    \begin{tabular}{|c|c|c|c|c|c|c|c|c|c|c|c|}
    \hline
Distr. & $D^{Tuk}$ & $D^{zon}$ & $D^{Mah}$ & $D^{Mah}_{MCD.75}$ & EM & regPCA1 & regPCA2 & $k$NN & RF & mean & oracle \\
    \hline
$t\,\infty$ & {\bf 1.656} & 1.754 & 1.853 & {\it 1.671} & 1.817 & 1.855 & 1.853 & 1.793 & 1.762 & 2.182 & 1.514 \\
 & {\bf (0.2523)} & (0.3142) & (0.4062) & {\it (0.2906)} & (0.3937) & (0.4158) & (0.4071) & (0.2974) & (0.282) & (0.3821) & (0.2121) \\ \cline{1-1}
$t\,10$ & {\bf 1.859} & 1.973 & 2.048 & {\it 1.865} & 2.027 & 2.05 & 2.044 & 1.995 & 1.968 & 2.45 & 1.677 \\
 & {\bf (0.3062)} & (0.4031) & (0.511) & {\it (0.3917)} & (0.4933) & (0.5069) & (0.5119) & (0.3861) & (0.3402) & (0.4778) & (0.279) \\ \cline{1-1}
$t\,5$ & {\bf 2.09} & 2.23 & 2.31 & {\it 2.109} & 2.267 & 2.348 & 2.31 & 2.255 & 2.233 & 2.749 & 1.91 \\
 & {\bf (0.4275)} & (0.5122) & (0.6217) & {\it (0.4841)} & (0.6006) & (0.6504) & (0.6219) & (0.4543) & (0.4476) & (0.6089) & (0.3742) \\ \cline{1-1}
$t\,3$ & {\bf 2.507} & 2.697 & 2.772 & {\it 2.541} & 2.737 & 2.791 & 2.779 & 2.707 & 2.699 & 3.32 & 2.239 \\
 & {\bf (0.6389)} & (0.7977) & (0.8516) & {\it (0.7133)} & (0.8243) & (0.9306) & (0.8495) & (0.6946) & (0.7254) & (0.964) & (0.5497) \\ \cline{1-1}
$t\,2$ & {\bf 3.462} & 3.68 & 3.733 & {\it 3.517} & 3.669 & 3.807 & 3.736 & 3.709 & 3.762 & 4.476 & 3.061 \\
 & {\bf (1.35)} & (1.577) & (1.6) & {\it (1.39)} & (1.589) & (1.648) & (1.601) & (1.413) & (1.444) & (1.794) & (1.136) \\ \cline{1-1}
$t\,1$ & {\bf 11.81} & 14.12 & 14.22 & {\it 12.34} & 13.78 & 13.73 & 14.31 & 12.58 & 13.64 & 15.73 & 10.37 \\
 & {\bf (9.738)} & (12.09) & (12.05) & {\it (9.631)} & (11.48) & (11.01) & (12.12) & (10.36) & (11.44) & (12.65) & (8.249) \\ \hline
    \end{tabular}
  }
  \caption{Median and MAD of the RMSEs of the imputation for $100$ points drawn from  elliptically symmetric Student-$t$ distributions with $\bmm_2$ and $\bmS_2$ contaminated with $15\%$ outliers, and $15\%$ of MCAR values on non-contaminated data, over 1000 repetitions.}
  \label{tab:StudentT100MCARoutl15}
  \end{center}
\end{table}

\begin{table}[!ht]
  \begin{center}
  {\tiny
    \begin{tabular}{|c|c|c|c|c|c|c|c|c|c|c|c|}
    \hline
Distr. & $D^{Tuk}$ & $D^{zon}$ & $D^{Mah}$ & $D^{Mah}_{MCD.75}$ & EM & regPCA1 & regPCA2 & $k$NN & RF & mean & oracle \\
    \hline
$t\,\infty$ & 1.464 & 1.454 & {\bf 1.447} & 1.453 & {\it 1.449} & 1.529 & {\bf 1.447} & 1.571 & 1.581 & 2.009 & 1.404 \\
 & (0.3694) & (0.3713) & {\bf (0.3595)} & (0.3663) & {\it (0.3593)} & (0.401) & {\bf (0.3594)} & (0.3892) & (0.3946) & (0.486) & (0.3399) \\ \cline{1-1}
$t\,10$ & 1.649 & 1.597 & {\bf 1.57} & {\it 1.572} & {\bf 1.57} & 1.665 & {\bf 1.57} & 1.755 & 1.754 & 2.2 & 1.529 \\
 & (0.4316) & (0.4285) & {\bf (0.4163)} & {\it (0.4203)} & {\bf (0.4206)} & (0.4502) & {\bf (0.4163)} & (0.4565) & (0.46) & (0.5737) & (0.4278) \\ \cline{1-1}
$t\,5$ & 1.816 & 1.799 & {\bf 1.757} & {\it 1.758} & {\bf 1.757} & 1.876 & {\bf 1.757} & 1.955 & 1.972 & 2.402 & 1.712 \\
 & (0.5134) & (0.5129) & {\bf (0.49)} & {\it(0.4991)} & {\bf (0.4899)} & (0.5499) & {\bf (0.4901)} & (0.555) & (0.5345) & (0.7318) & (0.4869) \\ \cline{1-1}
$t\,3$ & 2.213 & 2.184 & 2.147 & {\bf 2.101} & {\it 2.139} & 2.242 & 2.147 & 2.37 & 2.343 & 2.844 & 2.054 \\
 & (0.7882) & (0.8159) & (0.8016) & {\bf (0.7618)} & {\it (0.8)} & (0.7782) & (0.801) & (0.8563) & (0.8357) & (1.011) & (0.7649) \\ \cline{1-1}
$t\,2$ & 2.837 & 2.919 & 2.813 & {\bf 2.68} & {\it 2.8} & 2.911 & 2.813 & 3.03 & 2.99 & 3.578 & 2.529 \\
 & (1.249) & (1.342) & (1.309) & {\bf (1.196)} & {\it (1.287)} & (1.311) & (1.31) & (1.325) & (1.331) & (1.554) & (1.133) \\ \cline{1-1}
$t\,1$ & {\bf 7.806} & 8.718 & 8.911 & 8.286 & 8.9 & 9.118 & 8.935 & {\it 8.135} & 8.138 & 10.99 & 6.367 \\
 & {\bf (6.351)} & (7.135) & (7.127) & (6.602) & (7.124) & (7.334) & (7.137) & {\it (6.605)} & (6.563) & (8.952) & (5.12) \\ \hline
    \end{tabular}
  }
  \caption{Median and MAD of the RMSEs of the imputation for a sample of $100$ points drawn from  elliptically symmetric Student-$t$ distributions, with  $\bmm_2$ and $\bmS_2$  having $5\%$ of  MCAR values, over 1000 repetitions.}
  \label{tab:StudentT100MCAR05}
  \end{center}
\end{table}

\begin{table}[!ht]
  \begin{center}
  {\tiny
    \begin{tabular}{|c|c|c|c|c|c|c|c|c|c|c|c|}
    \hline
Distr. & $D^{Tuk}$ & $D^{zon}$ & $D^{Mah}$ & $D^{Mah}_{MCD.75}$ & EM & regPCA1 & regPCA2 & $k$NN & RF & mean & oracle \\
    \hline
$t\,\infty$ & {\bf 1.552} & 1.613 & 1.709 & {\it 1.553} & 1.701 & 1.769 & 1.709 & 1.695 & 1.603 & 2.167 & 1.406 \\
 & {\bf (0.3693)} & (0.4107) & (0.4867) & {\it (0.4379)} & (0.4788) & (0.5248) & (0.4877) & (0.407) & (0.3924) & (0.5981) & (0.3171) \\ \cline{1-1}
$t\,10$ & {\bf 1.706} & 1.778 & 1.874 & {\it 1.73} & 1.861 & 1.906 & 1.875 & 1.884 & 1.823 & 2.398 & 1.564 \\
 & {\bf (0.4415)} & (0.5106) & (0.6104) & {\it (0.4912)} & (0.6032) & (0.6053) & (0.6111) & (0.5182) & (0.4797) & (0.7059) & (0.3938) \\ \cline{1-1}
$t\,5$ & {\bf 1.868} & 1.951 & 2.038 & {\it 1.877} & 2.027 & 2.172 & 2.039 & 2.077 & 1.995 & 2.57 & 1.698 \\
 & {\bf (0.5565)} & (0.5843) & (0.6859) & {\it (0.5679)} & (0.6806) & (0.7747) & (0.6819) & (0.6256) & (0.6102) & (0.8625) & (0.491) \\ \cline{1-1}
$t\,3$ & {\it 2.243} & 2.348 & 2.421 & {\bf 2.226} & 2.42 & 2.525 & 2.421 & 2.429 & 2.392 & 3.05 & 2.016 \\
 & {\it (0.8064)} & (0.8694) & (0.9166) & {\bf (0.8258)} & (0.9345) & (1.019) & (0.9237) & (0.8484) & (0.8521) & (1.171) & (0.7047) \\ \cline{1-1}
$t\,2$ & {\bf 2.902} & 3.032 & 3.183 & {\it 2.933} & 3.163 & 3.196 & 3.188 & 3.142 & 3.071 & 4.073 & 2.55 \\
 & {\bf (1.375)} & (1.498) & (1.566) & {\it (1.421)} & (1.558) & (1.565) & (1.582) & (1.472) & (1.43) & (2.007) & (1.129) \\ \cline{1-1}
$t\,1$ & {\bf 7.464} & 8.487 & 8.531 & 8.334 & 8.5 & 8.675 & 8.541 & {\it 7.958} & 8.1 & 10.82 & 6.245 \\
 & {\bf (5.916)} & (6.869) & (7.081) & (6.867) & (6.988) & (7.261) & (7.117) & {\it (6.509)} & (6.922) & (8.802) & (4.874) \\ \hline
    \end{tabular}
  }
  \caption{Median and MAD of the RMSEs of the imputation for $100$ points drawn from  elliptically symmetric Student-$t$ distributions with $\bmm_2$ and $\bmS_2$ contaminated with $15\%$ of outliers, with $5\%$  MCAR values on non-contaminated data, over 1000 repetitions.}
  \label{tab:StudentT100MCARoutl05}
  \end{center}
\end{table}


\bigskip

\section{Proofs}

\noindent{\bf Proof of Theorem~1:} \indent\\
Due to the fact that $D_{n,\alpha}(\bmX)\xrightarrow[n\rightarrow\infty]{a.s.}D_{\alpha}(X)$, in what follows we focus on the population version only. For $X\sim\mathcal{E}_d({\boldsymbol\mu_X},{\boldsymbol\Sigma}_X,F_R)$ allow the transform $X\mapsto Z=\boldsymbol{R}\boldsymbol{\Sigma}^{-1/2}(X - \boldsymbol{\mu})$, with $\boldsymbol{R}$ being a rotation operator such that w.l.o.g. $\bmx\mapsto\bmz$, such that missing values still constitute a $|miss(\bmx)|$-dimensional affine space parallel to missing coordinates' axes. Since contours $D_\alpha(Z)$ are concentric spheres centered at the origin, $D_\alpha^*(Z)$ in~(3) is of the form $\{\bmv\,|\,\bmv = \bmz^\prime + \beta\bmr\,,\,\beta\ge0\}$ with $\bmz^\prime_{obs(\bmz)} = \bmz_{obs}$ and $\bmz^\prime_{miss(\bmz)}=\boldsymbol{0}_{|miss(\bmx)|}$, and $\bmr\in\mathcal{S}^{|miss(\bmx)|-1}$, a unit sphere in the linear span of $miss(\bmz)$. Because of the fact that $P\bigl(\{\bmx\in\mathbb{R}^d\,|\,D(\bmx|X)=\alpha\}\bigr)=0$, $\beta=0$ almost surely and thus $\bmz$ is imputed with $\bmz^\prime = \boldsymbol{R}\boldsymbol{\Sigma}^{-1/2}(\bmy - \boldsymbol{\mu})$.\hfill$\Box$

\bigskip

\noindent{\bf Proof of Theorem~2:} \indent\\
(The challenge here is that the resulting distribution is not elliptical.)

For $X\sim\mathcal{E}_d({\boldsymbol\mu_X},{\boldsymbol\Sigma}_X,F_R)$ allow the transform $X\mapsto Z=\boldsymbol{R}\boldsymbol{\Sigma}^{-1/2}(X - \boldsymbol{\mu})$, with $\boldsymbol{R}$ being a rotation operator such that w.l.o.g. $\bmx\mapsto\bmz$, such that $miss(\bmz)=1$. ($Z$ has spherical density contours and missing values are in the first coordinate only.)

Let $Z^\prime = \bigl(0, (Z^{\prime\prime})^\top\bigr)^\top$ with $Z^{\prime\prime}\sim\mathcal{E}_{d-1}({\boldsymbol{0}},{\boldsymbol{I}},F_R)$, where $\boldsymbol{I}$ is the diagonal matrix. Consider a random vector $U\sim(1-p)Z + pZ^{\prime}$ which is a mixture of $d$- and $(d-1)$-dimensional spherical distributions. $Z^{\prime}$ corresponds to the imputed missing values---let us now show that this is true. Due to the fact that $D_{n,\alpha}(\bmU)\xrightarrow[n\rightarrow\infty]{a.s.}D_{\alpha}(U)$, in what follows we focus on the population version only. Missing values constitute one-dimensional affine subspaces parallel to the first coordinate. Thus, due to the affine invariance property (P1 in Definition~2), $D_\alpha(U)\cap\{\bmu\in\mathbb{R}^d\,|\,\bmu_1\ge 0\}$ $=$ $D_\alpha(U)\cap\{\bmu\in\mathbb{R}^d\,|\,\bmu_1\le 0\}\times(-1,0,\ldots,0)^\top$. To see this, it  suffices to note that the symmetric reflection of $U$ w.r.t. the  linear space normal to $(1,0,\ldots,0)^\top$ equals $U$. Now, for $\lambda\in\mathbb{R}$ let $\bmu=(\lambda,\bmu_2,\ldots,\bmu_d)^\top$ be this one-dimensional affine subspace of missingness for a point. In~(3), $\text{ave}\bigl(D_\alpha(U)\cap\bmu\bigr)$ $=$ $\text{ave}\bigl(D_\alpha(U)\cap\{\bmv\in\mathbb{R}^d\,|\,\bmv_1>0\}\cap\bmu \cup D_\alpha(U)\cap\{\bmv\in\mathbb{R}^d\,|\,\bmv_1<0\}\cap\bmu \cup D_\alpha(U)\cap\{\bmv\in\mathbb{R}^d\,|\,\bmv_1=0\}\cap\bmu\bigr)$ $=$ $D_\alpha(U)\cap\{\bmv\in\mathbb{R}^d\,|\,\bmv_1=0\}\cap\bmu$ $=$ $(0,\bmu_2,\ldots,\bmu_d)$ $=$ $\boldsymbol{R}\boldsymbol{\Sigma}^{-1/2}(\bmy - \boldsymbol{\mu})$ (with the obvious correspondence between $\bmu$ and $\bmy$).\hfill$\Box$

\bigskip

\noindent{\bf Proof of Corollary~1:} \indent\\
(P1)--(P5) are obviously satisfied for the Tukey, zonoid and Mahalanobis depths. In Theorem~1, $D_{n,\alpha}(\bmX)\xrightarrow[n\rightarrow\infty]{a.s.}D_{\alpha}(X)$ is clearly satisfied for the Mahalanobis depth, following Corollary~3.11 by \cite{Mosler02} for the zonoid depth, and by Theorem~4.2 in \cite{ZuoS00b} for the Tukey depth. In Theorem~2, the same logic holds for the Mahalanobis and zonoid depths, but not for the Tukey depth as $Z$ is not elliptical. Using techniques similar those in the proof of Theorem~3.4 in \cite{ZuoS00b}, one can show that $P\bigl(\{\bmx\in\mathbb{R}^d\,|\,D(\bmx|Z)=\alpha\}\bigr)=0$, from which, together with the vanishing at infinity property (P4) and $\sup_{\bmx\in\mathbb{R}^d}|D_n(\bmx|\bmZ) - D(\bmx|Z)|\xrightarrow[n\rightarrow\infty]{a.s.}0$ \citep[see][]{DonohoG92}, it follows that $D_{n,\alpha}(\bmZ)\xrightarrow[n\rightarrow\infty]{a.s.}D_{\alpha}(Z)$.\hfill$\Box$

\bigskip

\noindent{\bf Proof of Proposition~1:} w.l.o.g. we restrict ourselves to the case $i=1$. Let $\bmZ$ be $\bmX$ transformed in such a way that it is an $n\times d$ matrix with $\bmm_\bmZ={\boldsymbol 0}$ and $\bmz_{1,miss(1)}=\bmS_{\bmZ\,miss(1),obs(1)}\bmS_{\bmZ\,obs(1),obs(1)}^{-1}\bmz_{1,obs(1)}$. Denote the argument $\bma=(0,\ldots,0,\bmy^\top)^\top\in\mathbb{R}^d$. Replacing $\bmz_1$ with $\bmz_1 + \bma$ and subtracting the column-wise average $\frac{\bma}{n}$ from each row gives the covariance matrix estimate:
\begin{align*}
  n\bmS_\bmZ(\bmy) &= \bmZ^\top\bmZ - \bmz_1\bmz_1^\top+ (\bmz_1 + \bma)(\bmz_1 + \bma)^\top - \frac{1}{n}\bma\bma^\top \\
  &= \bmZ^\top\bmZ + 2\bmz_1\bma^\top + \frac{n - 1}{n}\bma\bma^\top\,.
\end{align*}
Since $\bmz_1^\top(\bmZ^\top\bmZ)^{-1}\bma=0$ due to Mahalanobis orthogonality, by simple algebra for the determinant, one obtains:
\begin{align*}
	n^d|\bmS_\bmZ(\bmy)| &= \Bigl| \bmZ^\top\bmZ + \sqrt{2}\bmz_1\bma^\top\sqrt{2} + \sqrt{\frac{n - 1}{n}}\bma\bma^\top\sqrt{\frac{n - 1}{n}} \Bigr| \\
	&= \bigl| \bmZ^\top\bmZ + \sqrt{2}\bmz_1\bma^\top\sqrt{2} \bigr| \Bigl( 1 + \sqrt{\frac{n - 1}{n}}\bma^\top\bigl(\bmZ^\top\bmZ + \sqrt{2}\bmz_1\bma^\top\sqrt{2}\bigr)^{-1}\bma\sqrt{\frac{n - 1}{n}} \Bigr) \\
	&= |\bmZ^\top\bmZ|\bigl(1 + \sqrt{2}\bma^\top(\bmZ^\top\bmZ)^{-1}\bmz_1\sqrt{2}\bigr)\Bigr( 1 + \sqrt{\frac{n - 1}{n}}\bma^\top\bigl(\bmZ^\top\bmZ\bigr)^{-1}\bma\sqrt{\frac{n - 1}{n}} - \\
	&- \frac{\sqrt{\frac{n - 1}{n}}\bma^\top(\bmZ^\top\bmZ)^{-1}\bmz_1\sqrt{2}\cdot\sqrt{2}\bma^\top(\bmZ^\top\bmZ)^{-1}\bma\sqrt{\frac{n - 1}{n}}}{1 + \sqrt{2}\bma^\top(\bmZ^\top\bmZ)^{-1}\bmz_1\sqrt{2}} \Bigr) \\
	& = |\bmZ^\top\bmZ|\bigl(1 + \frac{n - 1}{n}\bma(\bmZ^\top\bmZ)^{-1}\bma\bigr)\,.
\end{align*}
Thus $|\bmS_\bmZ(\bmy)|$ is a quadratic function of $\bmy$, which is clearly minimized in $\bmy = (0,\ldots,0)^\top$.
\hfill$\Box$

\bigskip

\noindent{\bf Proof of Theorem~3:} The first point can be checked by elementary algebra. The second point follows from the coordinate-wise application of Proposition~1. For the third point, it suffices to prove the single-output regression case. The regularized PCA algorithm will converge if
\begin{align*}
\bmy_{id} = \sum_{s=1}^d u_{is}\sqrt{\lambda_s}v_{ds} = \sum_{s=1}^d u_{is}(\sqrt{\lambda_s}-\frac{\sigma^2}{\sqrt{\lambda_s}})v_{ds}
\end{align*}
for any $\sigma^2\le\lambda_d$. W.l.o.g. we prove that
\begin{align*}
\bmy_d=\bmS_{d\,(1,...,d-1)}\bmS^{-1}_{(1,...,d-1)\,(1,...,d-1)}\bmy_{(1,...,d-1)}\,\Longleftrightarrow\,\sum_{i=1}^d\frac{u_iv_{di}}{\sqrt{\lambda_i}}=0,
 \end{align*}
denoting $\bmS(\bmY)$ simply $\bmS$ for the centered $\bmY$, and an arbitrary point $\bmy$. Using matrix algebra,
\begin{align*}
\bmy_d &=\bmS_{d\,(1,...,d-1)}\bmS^{-1}_{(1,...,d-1)\,(1,...,d-1)}\bmy_{(1,...,d-1)} = -\bigl((\bmS^{-1})_{dd}\bigr)^{-1}(\bmS^{-1})_{d\,(1,...,d-1)}\bmy_{(1,...,d-1)}, \\
\sum_{i=1}^d u_i\sqrt{\lambda_i}v_{di} &= - \Bigl(\sum_{i=1}^d\frac{v_{di}^2}{\lambda_i}\Bigr)^{-1} \Bigl(\sum_{i=1}^d\frac{v_{di}v_{1i}}{\lambda_i},\sum_{i=1}^d\frac{v_{di}v_{2i}}{\lambda_i},...,\sum_{i=1}^d\frac{v_{di}v_{(d-1)\,i}}{\lambda_i}\Bigr)\times \\
&\times \Bigl(\sum_{i=1}^d u_i\sqrt{\lambda_i}v_{1i},\sum_{i=1}^d u_i\sqrt{\lambda_i}v_{2i},...,\sum_{i=1}^d u_i\sqrt{\lambda_i}v_{(d-1)\,i}\Bigr)^\top.
\end{align*}
After reordering the terms, one obtains
\begin{align*}
\sum_{i=1}^d u_i\sqrt{\lambda_i} \sum_{j=1}^d\frac{v_{dj}}{\lambda_j} \sum_{k=1}^d v_{ki}v_{kj}=0.
\end{align*}
Due to the orthogonality of ${\boldsymbol V}$, $d^2-d$ terms from the two outer sum signs are zero. Gathering non-zero terms, i.e., those with $i=j$ only, we have that
\begin{align*}
\sum_{i=1}^d u_i\sqrt{\lambda_i}\frac{v_{di}}{\lambda_i}=\sum_{i=1}^d\frac{u_iv_{di}}{\sqrt{\lambda_i}}=0.
\end{align*}
\hfill$\Box$

\bigskip

\begin{figure}[!ht]\centering
    \includegraphics[keepaspectratio=true,scale=.785]{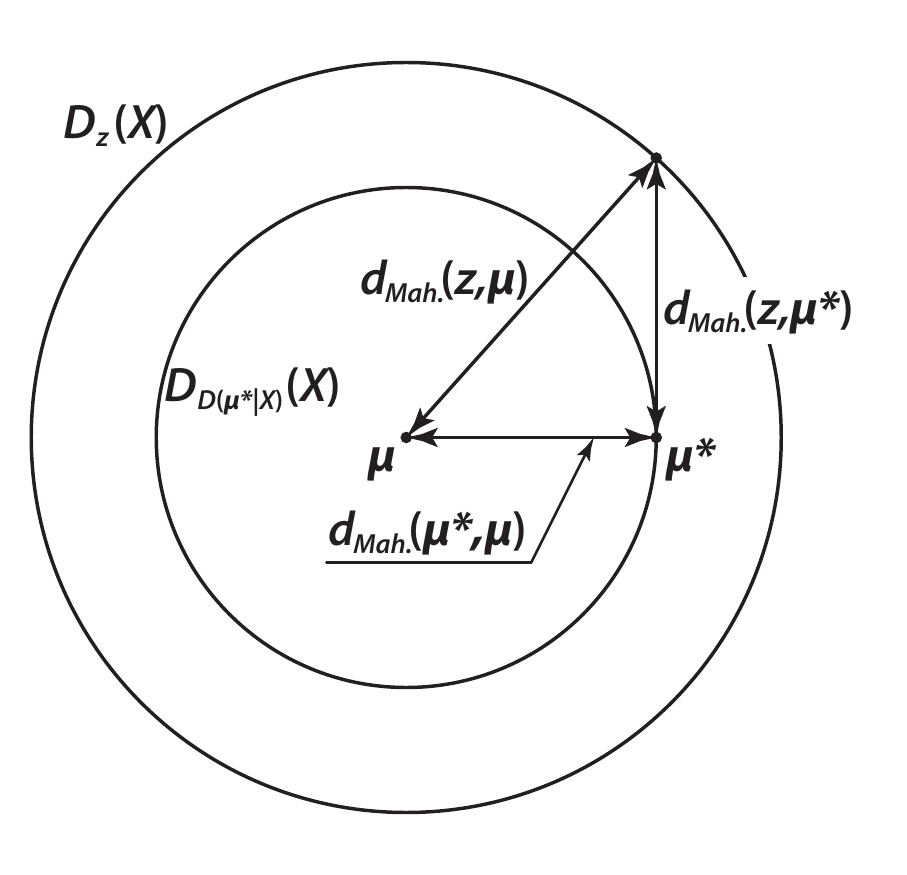}\includegraphics[keepaspectratio=true,scale=.785]{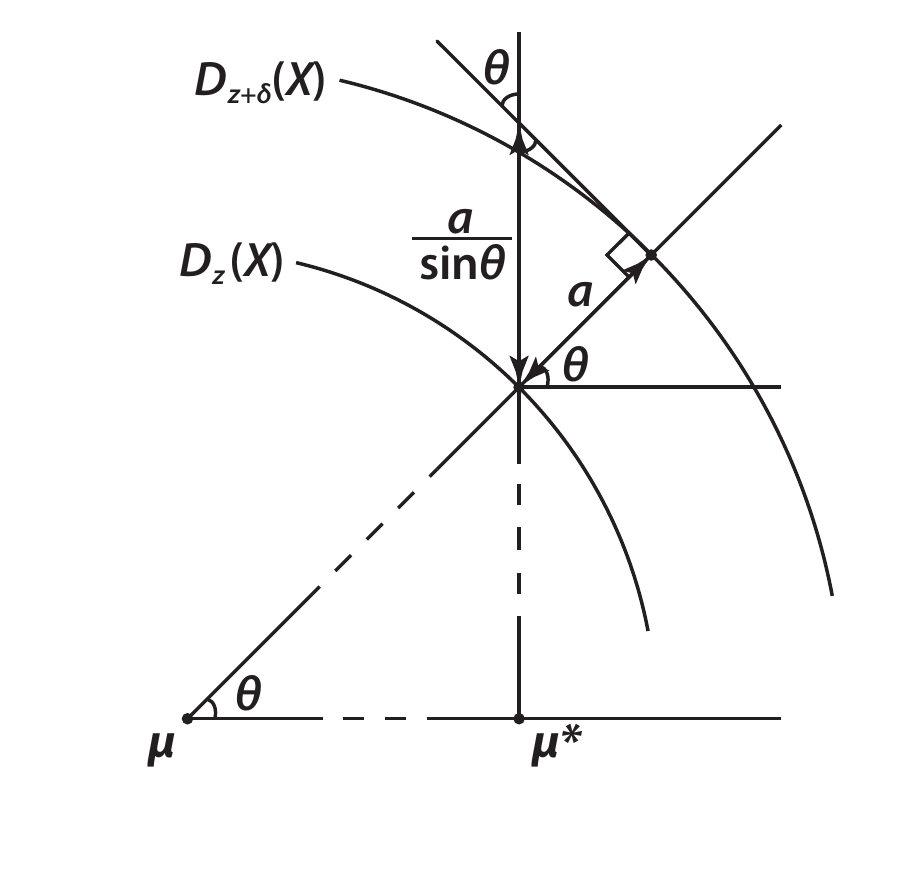}
	\caption{Illustration of the derivation of (4).}
	\label{fig:ellipseAngle}
\end{figure}

\noindent {\bf Derivation of (4):} The integrated quantity is the conditional depth density that can be obtained from the joint one by the volume transformation (denoting $d_{M}\bigl(z,\bmm\bigr)$ the Mahalanobis distance between a point of depth $z$ and $\bmm$):
\begin{align*}
f_{D((X|X_{obs}=\bmx_{obs})|X)}(z) &= f_{D(X|X)}(z)\cdot C\cdot T_{down}\bigl(d_{M}(z,\bmm)\bigr)\cdot T_{up}\bigl(d_{M}(z,\bmm^*)\bigr)\times \\
&\times T_{angle}\bigl(d_{M}(z,\bmm),d_{M}(z,\bmm^*)\bigr)\,.
\end{align*}
Any constant $C$ is ignored as it is unimportant when drawing.
The three terms below correspond to descaling the density to dimension one (downscaling), re-scaling it to the dimension of the missing values (upscaling), and the linear transformation from dimension $d$ to dimension $|miss|=$number of missing coordinates of a point (angle transformation): 
\begin{align*}
T_{down}\bigl(d_{M}(z,\bmm)\bigr) &= d_{M}^{1-d}(z,\bmm) = \frac{1}{d_{M}^{d-1}(z,\bmm)}\,. \\
T_{up}\bigl(d_{M}(z,\bmm^*)\bigr) &= d_{M}^{|miss(\bmx)|-1}(z,\bmm^*) \\
&= \Bigl(\sqrt{d^2_{M}(z,\bmm) - d^2_{M}\bigl(D(\bmm^*|X),\bmm\bigl)}\Bigr)^{|miss(\bmx)|-1}\,. \\
T_{angle}\bigl(d_{M}(z,\bmm),d_{M}(z,\bmm^*)\bigr) &= \frac{1}{\sin\theta} = \frac{1}{\frac{d_{M}(z,\bmm^*)}{d_{M}(z,\bmm)}} \\
&= \frac{d_{M}(z,\bmm)}{\sqrt{d^2_{M}(z,\bmm) - d^2_{M}\bigl(D(\bmm^*|X),\bmm\bigl)}}\,.
\end{align*}
$T_{down}$ and $T_{up}$ are illustrated in Figure~\ref{fig:ellipseAngle} (left); for $T_{angle}$ see Figure~\ref{fig:ellipseAngle} (right). Setting $d_{M}(z,\bmm) = d_{M}(z)$ to shorten notation gives~(4).

\end{document}